\documentclass[12pt]{article}
\pdfoutput=1
\usepackage{jheppub}
\usepackage{verbatim}
\usepackage{amsmath}
\usepackage{amssymb}
\usepackage{amsthm}
\usepackage{slashed}
\usepackage{amsfonts}
\usepackage[utf8]{inputenc}
\usepackage[normalem]{ulem}

\def\be{\begin{equation}}
\def\ee{\end{equation}}
\def\ba#1\ea{\begin{align}#1\end{align}}
\def\bg#1\eg{\begin{gather}#1\end{gather}}
\def\bm#1\em{\begin{multline}#1\end{multline}}
\def\bmd#1\emd{\begin{multlined}#1\end{multlined}}

\def\a{\alpha}
\def\b{\beta}
\def\c{\chi}

\def\g{\gamma}

\def\m{\mu}

\def\p{\phi}

\def\r{\rho}

\def\S{\Sigma}

\def\y{\psi}

\def\la{\label}

\def\re{\ref}
\def\er{\eqref}

\def\fr{\frac}

\def\pa{\partial}

\def\wtd{\widetilde}
\def\ol{\overline}

\def\eq{\equiv}

\def\({\left(}
\def\){\right)}
\def\[{\left[}
\def\]{\right]}
\def\<{\langle}
\def\>{\rangle}

\def\bZ{{\mathbb Z}}

\def\cL{{\mathcal L}}

\def\cR{{\mathcal R}}

\newcommand{\bfig}{\begin{figure}\begin{center}}
\newcommand{\efig}{\end{center}\end{figure}}
\newcommand{\bi}{\begin{itemize}}
\newcommand{\ei}{\end{itemize}}

\newcommand{\lan}{\langle}
\newcommand{\ran}{\rangle}
\newcommand{\Tr}{\mathrm{Tr}}

\newcommand{\wt}{\widetilde}

\newcommand{\LL}{\mathcal{L}}

\newcommand{\Hra}{\mathcal{H}_{r_\alpha}}
\newcommand{\Hrba}{\mathcal{H}_{\ol{r}_\alpha}}
\newcommand{\Hc}{\mathcal{H}_{code}}
\newcommand{\HR}{\mathcal{H}_R}
\newcommand{\HRb}{\mathcal{H}_{\ol{R}}}

\theoremstyle{definition}

\begin{document}

\title{Flat entanglement spectra in fixed-area states of quantum gravity}
\author[a]{Xi Dong,}
\author[b]{Daniel Harlow,}
\author[a]{and Donald Marolf}
\affiliation[a]{Department of Physics, University of California, Santa Barbara, CA 93106, USA}
\affiliation[b]{Center for Theoretical Physics, Massachusetts Institute of Technology, Cambridge, MA 02139, USA}
\emailAdd{xidong@ucsb.edu}
\emailAdd{harlow@mit.edu}
\emailAdd{marolf@ucsb.edu}

\abstract{We use the Einstein-Hilbert gravitational path integral to investigate gravitational entanglement at leading order $O(1/G)$. We argue that semiclassical states prepared by a Euclidean path integral have the property that projecting them onto a subspace in which the Ryu-Takayanagi or Hubeny-Rangamani-Takayanagi surface has definite area gives a state with a flat entanglement spectrum at this order in gravitational perturbation theory. This means that the reduced density matrix can be approximated as proportional to the identity to the extent that its Renyi entropies $S_n$ are independent of $n$ at this order. The $n$-dependence of $S_n$ in more general states then arises from sums over the RT/HRT-area, which are generally dominated by different values of this area for each $n$. This provides a simple picture of gravitational entanglement, bolsters the connection between holographic systems and tensor network models, clarifies the bulk interpretation of algebraic centers which arise in the quantum error-correcting description of holography, and strengthens the connection between bulk and boundary modular Hamiltonians described by Jafferis, Lewkowycz, Maldacena, and Suh.}

\maketitle

\section{Introduction}
The study of entanglement in gravitational systems has a long history, going back to \cite{Sorkin:2014kta,Bombelli:1986rw,Srednicki:1993im,Frolov:1993ym}.  In recent years it has been given new life within the relatively precise context of AdS/CFT \cite{Maldacena:2001kr,Ryu:2006bv,Ryu:2006ef,Hubeny:2007xt,Swingle:2009bg,VanRaamsdonk:2010pw,Bianchi:2012ev,Lewkowycz:2013nqa,Maldacena:2013xja,Balasubramanian:2014hda,Marolf:2015vma,Freedman:2016zud}.  The primary driver of this resurgence of interest has been the Ryu-Takayanagi (RT) formula, which says that at leading order in the semiclassical expansion in $G$ the von Neumann entropy on a boundary spatial subregion $R$ of any semiclassical state $\rho$ is given by
\be\label{leadingRT}
S(\rho_R)=\frac{A[\gamma_R]}{4G},
\ee
where $\gamma_R$ is the Hubeny-Rangamani-Takayanagi (HRT) surface in the bulk associated to that subregion and $A[\gamma_R]$ is its area in that state \cite{Ryu:2006bv,Ryu:2006ef,Hubeny:2007xt}. More recently, the nature of entanglement in AdS/CFT was clarified considerably by the observation that the holographic mapping from the bulk to the boundary has the structure of a quantum error-correcting code \cite{Almheiri:2014lwa}.  Quantum error-correcting codes are protocols which store quantum information nonlocally in the entanglement between many local degrees of freedom, in such a way that the stored information is protected from errors acting on small numbers of those local degrees of freedom: in \cite{Almheiri:2014lwa} it was observed that this is precisely what is required to explain the emergence of the radial direction in AdS/CFT.  In particular in \cite{Jafferis:2015del,Dong:2016eik,Harlow:2016vwg} a close connection between the RT formula and quantum error correction was developed, with \cite{Harlow:2016vwg} showing that a rather general family of quantum error-correcting codes all obey an RT-like formula.

Throughout these developments, a supporting role has been played by Renyi entropies, which for any quantum state $\rho$ are defined by
\be\label{sndef}
S_n(\rho)\equiv -\frac{1}{n-1}\log\Tr(\rho^n).
\ee
Typically these are well-defined for $n\geq 1$, with $S_1(\rho)$ being equivalent to the von Neumann entropy $-\Tr\rho \log \rho$.  Indeed the most prominent appearance of Renyi entropy is in the use of the replica trick to compute von Neumann entropy via the limit $n\to 1$ \cite{Holzhey:1994we,Calabrese:2004eu,Headrick:2010zt,Lewkowycz:2013nqa}.  Renyi entropies are also interesting objects on their own however, with their $n$-dependence allowing them to probe more information about a quantum state than the von Neumann entropy does: indeed in principle we should be able to extract the full spectrum of $\rho$ from a careful study of the Renyi entropies (see \cite{Perlmutter:2013gua,Perlmutter:2013paa,Dong:2016fnf,Dong:2018lsk} for recent work focused specifically on Renyi entropies).  It therefore is natural to ask if quantum error correction gives any useful perspective on Renyi entropy.

\bfig
\includegraphics[height=7cm]{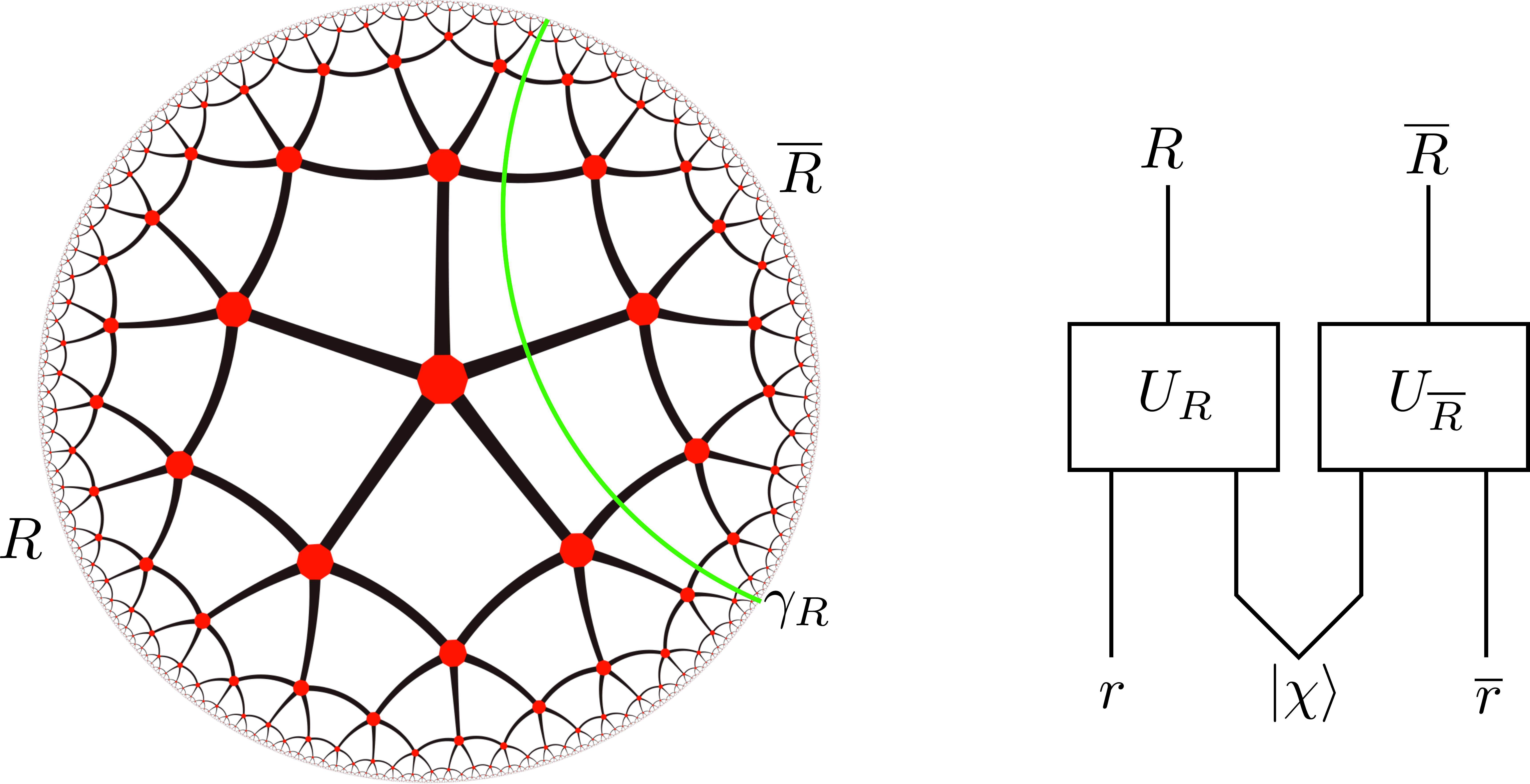}
\caption{The encoding circuit for a tensor network holographic code (figure borrowed from \cite{Harlow:2018fse}).  $r$/$\ol{r}$ denotes bulk degrees of freedom at the red dots to the left/right of the green HRT surface $\gamma_R$, $|\chi\ran$ is a tensor product of a set of EPR pairs on each link cut by $\gamma_R$, and $U_R$/$U_{\ol{R}}$ are unitary transformations generated by the pieces of the network to the left/right of $\gamma_R$.}\label{cutnetworkfig}
\efig
One way to begin to address this question is to ask how Renyi entropy behaves in simple tensor network models of holographic codes which have so far been constructed \cite{Pastawski:2015qua,Hayden:2016cfa}, but here a surprise is in order.  These models are (to a good approximation in the second case) examples of what \cite{Harlow:2016vwg} called subsystem codes with complementary recovery, which in practice means that they are encoded via a quantum circuit of the type illustrated in figure \ref{cutnetworkfig}.\footnote{Readers for whom this is unfamiliar may wish to consult \cite{Harlow:2018fse} for more background on these ideas, which we will also review in some more detail in section \ref{qecsec} below.}  The quantum error correction interpretation of the RT formula proposed in \cite{Harlow:2016vwg} follows from the circuit diagram in figure \ref{cutnetworkfig}: for any state $\rho_{r\ol{r}}$ we feed into the $r\ol{r}$ indices of the circuit, we have
\be\label{networkRT}
S(\rho_R)=S(\chi_R)+S(\rho_r),
\ee
where $\chi_R$ is the restriction of $|\chi\ran$ to the subfactor of $R$ which does not contain $r$.  $S(\chi_R)$ thus receives a contribution from each link which is cut by $\gamma_R$, and is therefore proportional to its area, so we can identify the first term in \eqref{networkRT} as giving rise to the leading-order RT formula \eqref{leadingRT}.  $S(\rho_r)$ is the von Neumann entropy of the bulk fields in the entanglement wedge of $R$, and thus gives the $O(G^0)$ contribution to the quantum version of the Ryu-Takayanagi formula proposed in \cite{Faulkner:2013ana}.  For our purposes here the main point however is that we can also use this encoding circuit diagram to compute the Renyi entropies of any bulk state we feed into the $r\ol{r}$ legs of the circuit.  At leading order in $G$ these Renyi entropies will again be dominated by the contribution from $|\chi\ran$, but since $|\chi\ran$ consists entirely of maximally mixed EPR pairs, the eigenvalues of its restriction $\chi_R$ to $R$ are all equal and the Renyi entropy $S_n(\rho_R)$ is thus independent of $n$ at leading order!  This certainly is not what is predicted by general relativity in the bulk (see e.g.~\cite{Dong:2016fnf}), and understanding what feature of holography is therefore missing has been one of the interesting open problems in holographic error correction.

The first guess for how to rectify this discrepancy is to change the tensors in the network.  This however will not change the $n$-independence of the Renyi entropies, since the index contradictions will still correspond to inserting maximally mixed states on each link.  We can try to improve this by changing the rules for doing the index contractions to include the insertion of a non-maximally mixed state \cite{Hayden:2016cfa}, but this seems unlikely to capture the full $n$-dependence expected from holography, especially e.g.\ in the presence of multiple intervals.  Our main goal in this paper is to instead argue that in a certain sense the simple holographic networks have it right: in the gravitational states which they are most analogous to, the Renyi entropies \textit{are} independent of $n$ to leading order in $G$!  Indeed we argue that holographic tensor networks of the type shown in figure \ref{cutnetworkfig} should best be understood as states where the area of the HRT surface $\gamma_R$ has been projected onto a definite value, killing the fluctuations which would usually be present in a good semiclassical state generated by a path integral.  In other words, the leading-order $n$-dependence of the Renyi entropies for such semiclassical states arises entirely from those fluctuations.

To reproduce this $n$-dependence using tensor networks, we therefore need to make the network geometry dynamical in some sense. This conclusion will be no surprise to experts, and indeed has already been discussed in \cite{Han:2017uco}\footnote{The simplest way to introduce dynamics to a tensor network is to include new degrees of freedom on the links which allow the tensor we sew in to change dynamically \cite{Donnelly:2016qqt}, which roughly speaking is what the ``center'' degrees of freedom we describe below in the gravitational picture are doing.}; see also \cite{Cantcheff:2014zma} for related discussion of HRT-area-eigenstates and their superpositions.  After all making the geometry of the HRT surface dynamical was the difference between an incorrect first attempt \cite{Fursaev:2006ih} to derive the RT formula, which just replicated the bulk geometry, and the later correct version \cite{Lewkowycz:2013nqa}, which allowed the geometry to backreact as needed to solve the equations of motion at the HRT surface \cite{Headrick:2010zt}.  Our contribution is to show more clearly how those two approaches are related, and that in particular there is a question for which the ``wrong'' replicated saddle point is actually the right answer.

The majority of this paper will focus on establishing the $n$-independence of the Renyi entropies of fixed-area states in gravity, but in section \ref{qecsec} we will return to quantum error correction to interpret our gravitational results.  In section \ref{jlmssec} we will then explain how our results imply a strengthening of the JLMS relation \cite{Jafferis:2015del} between bulk and boundary modular Hamiltonians.  The idea is that that relation holds also ``in the exponent'' as a statement about the bulk and boundary modular flow operators.  These operators are used to define the modular flow operation, so understanding them better may be of use in implementing the proposal of \cite{Faulkner:2017vdd} to use modular flow as an explicit bulk algorithm for entanglement wedge reconstruction.

\section{Cutting gravitational path integrals to compute boundary Renyi entropy}
The basic idea of this section is to cut and paste gravitational path integrals in a way that enables us to compute the boundary Renyi entropy of a state which has been projected to a definite area of the HRT surface. Doing so however requires us to understand how to describe gravity in a subregion.  This question has been studied in considerable detail at the classical level in \cite{Donnelly:2016auv} (see also \cite{Marolf:1994cp,Jafferis:2015del,Donnelly:2017jcd,Donnelly:2018nbv,Speranza:2017gxd,Kirklin:2018gcl,Camps:2018wjf,Giddings:2019hjc}
for related work); we now review it from a slightly different perspective which is more conducive to seeing the connection to quantum error correction we develop in section \ref{qecsec}.

\subsection{A phase space for gravity in the entanglement wedge}\label{subregionsec}
In any mechanical system, phase space is defined as the set of distinct initial conditions.  When the initial-value problem is well-posed, meaning that each initial condition leads to a unique classical solution, we can equivalently think of phase space as the set of classical solutions.  In the presence of gauge symmetries the initial-value problem is \textit{not} well-posed for the degrees of freedom appearing in the action, but it will be well-posed once we quotient the set of classical solutions by the set of gauge transformations: the phase space is then in one-to-one correspondence with these equivalence classes.

For example consider the Maxwell theory, with action
\be
S=-\frac{1}{2}\int_M F\wedge \star F.
\ee
Here $M$ is an arbitrary manifold, possibly with a codimension-one boundary where we need to specify boundary conditions.  For concreteness we will require that the pullback of $A$ to $\partial M$ vanishes, but other boundary conditions can be incorporated easily into our analysis.   We may define a ``pre-phase space''  consisting of all $A$ obeying these boundary conditions and also the equation of motion $d\star F=0$, and the physical phase space will then be the quotient of these by the set of gauge transformations $A'=A+d\epsilon$ such that $\epsilon|_{\partial M}=0$.  The pre-symplectic form on pre-phase space is given by
\be\label{presymp}
\Omega(\delta_1A,\delta_2A)=\int_\Sigma\left(\delta_1A\wedge \star d \delta_2 A-\delta_2 A\wedge \star d \delta_1 A\right),
\ee
where $\delta_1 A$ and $\delta_2 A$ are variations which  obey Maxwell's equation (they can be thought of as differentials on pre-phase space), and $\Sigma$ is any Cauchy slice of $M$ (it is easy to see that $\Omega$ is independent of the choice of $\Sigma$).  Note that if we take either of these variations to be a gauge transformation then $\Omega$ vanishes:
\begin{align}\nonumber
\Omega(d\epsilon,\delta A)&=\int_\Sigma d\epsilon\wedge \star d\delta A\\\nonumber
&=\int_{\partial\Sigma}\epsilon \star d\delta A\\
&=0.
\end{align}
In going from the first to the second line we have used that $\delta A$ obeys the equations of motion, while in going from the second to the third we have used that $\epsilon|_{\partial M}=0$.  Therefore the presymplectic form \eqref{presymp} is degenerate: the nondegenerate physical symplectic form is obtained once we quotient the set of $A$ by all gauge transformations obeying the boundary conditions.

The quotient from pre-phase space to phase space can lead to difficulties in trying to define subsystems of theories with gauge symmetries.  For example let $r$ be a subregion of $\Sigma$.\footnote{This is the first instance of a convention we will maintain throughout: $r$ denotes a spatial subregion which we will ultimately think of as being in the bulk of AdS/CFT, while $R$ denotes a boundary spatial subregion.} We can define a restricted pre-symplectic form
\be
\Omega_r(\delta_1A,\delta_2A)=\int_r\left(\delta_1A\wedge \star d \delta_2 A-\delta_2 A\wedge \star d \delta_1 A\right)
\ee
which we might hope to use in defining a dynamics of ``the degrees of freedom in $r$''. Unfortunately the same calculation that we just did for $\Omega$ now shows that $\Omega_r$ will \textit{not} be zero acting on variations which are pure gauge: instead we have
\be
\Omega_r(d\epsilon,\delta A)=\int_{\hat{\partial r}}\epsilon \star d\delta A,
\ee
where $\hat{\partial r}$ denotes the part of $\partial r$ which does not intersect $\partial \Sigma$.  This in general does not vanish.  There are various approaches to this problem which have been proposed in the literature.  In \cite{Donnelly:2011hn,Ghosh:2015iwa,Soni:2015yga,Donnelly:2016auv} the strategy is to promote the gauge transformations which do not vanish on $\hat{\partial r}$ (modulo the ones that do) to new physical degrees of freedom, which are sometimes called edge modes.  These degrees of freedom have no counterparts on the physical phase space on $\Sigma$, so the phase space one constructs this way is \textit{not} a submanifold of the original one.  The approach we will take instead is inspired by the algebraic approach of \cite{Casini:2013rba}, which we prefer since no unphysical degrees of freedom appear and the connection to quantum error correction is more manifest.\footnote{Our technique is also close to that in \cite{Jafferis:2015del}, but we differ on a few points.  Most importantly their analysis of the gravitational case might be read as implying that the restricted phase space only makes sense if the internal boundary $\hat{\partial r}$ is extremal (this claim was also recently made explicitly in \cite{Camps:2018wjf}), but our analysis makes it clear that there is no such requirement and any gauge-invariant choice of that boundary should work.}  Our strategy is instead to just fix the restriction of $A$ to $\hat{\partial r}$ to some definite configuration $A_{\hat{\partial r}}$, and then quotient only by gauge transformations in $r$ which vanish on $\partial r$: we will refer to the resulting phase space, on which $\Omega_r$ is a non-degenerate symplectic form, as $\mathcal{P}_r(A_{\hat{\partial r}})$.  Similarly on the complement $\ol{r}$ of $r$, we can define an analogous phase space $\mathcal{P}_{\ol{r}}(A_{\hat{\partial r}})$, where we have simplified notation by observing that $\hat{\partial r}=\hat{\partial \ol{r}}$.  We can then construct a gauge-invariant phase space on all of $\Sigma$ via
\be\label{phasedecomp}
\mathcal{P}=\coprod_{\alpha\in S}\left(\mathcal{P}_r(\alpha)\times \mathcal{P}_{\ol{r}}(\alpha)\right),
\ee
where $\coprod$ denotes disjoint union and $S$ denotes a set which contains a single representative of each gauge-equivalence class of the set of gauge field configurations on $\hat{\partial r}$.  Due to the presence of gauge symmetry, we see that it is not the product of a phase space for $r$ and a phase space for $\ol{r}$.  This phase space is not quite the one we started with, since it does not include the canonical conjugates of the $\alpha$ (the electric fields within $\hat{\partial r}$).  These two phase spaces however will lead to the same quantum theory: after quantization the decomposition \eqref{phasedecomp} of phase space becomes a Hilbert space decomposition
\be\label{hilbertdecomp}
\mathcal{H}=\oplus_{\alpha\in S}\left(\mathcal{H}_{r_\alpha}\otimes \mathcal{H}_{\ol{r}_\alpha}\right),
\ee
and the electric fields within $\hat{\partial r}$ reappear as operators which mix the different $\alpha$ sectors. We thus have not lost anything compared to what we would have gotten starting with the full theory on $\Sigma$.

The Hilbert space structure \eqref{hilbertdecomp} has an elegant interpretation via the theory of von Neumann algebras, which are subsets of the bounded operators on a Hilbert space that are closed under addition, multiplication, and hermitian conjugation, and which also contain all multiples of the identity.\footnote{If the Hilbert space it acts on is infinite-dimensional, a von Neumann algebra is further required to be closed in the weak operator topology.  The theorem we state momentarily applies also in infinite dimensions to von Neumann algebras which are direct sums of type I factors.}  A standard theorem (see e.g.\ the appendix of \cite{Harlow:2016vwg}) says that a von Neumann algebra $M$ acting on a finite-dimensional Hilbert space $\mathcal{H}$ always induces a decomposition of $\mathcal{H}$ of precisely the form \eqref{hilbertdecomp}, with all operators in either $M$ or its commutant $M'$ being block diagonal in the $\alpha$.  Moreover we can choose the tensor factorization within each block such that $M$ acts nontrivially only on $\Hra$ and $M'$ acts nontrivially only on $\Hrba$, and indeed any operator which is block diagonal and acts only on the $\Hra$ is in $M$ and any operator which is block diagonal and acts only on the $\Hrba$ is in $M'$.  The center $Z_M$ of $M$, which is also the center of $M'$, consists precisely of those block diagonal matrices which within each block are proportional to the identity on $\Hra\otimes\Hrba$.  In other words we can decompose the Hilbert space as in \eqref{hilbertdecomp} such that
\begin{align}\nonumber
M&=\oplus_\alpha\left(\LL(\Hra)\otimes I_{\ol{r}_\alpha}\right),\\
M'&=\oplus_\alpha\left(I_{r_\alpha}\otimes \LL(\Hrba)\right),\\\nonumber
Z_M&=\oplus_\alpha \lambda_\alpha I_{r_\alpha \ol{r}_\alpha},
\end{align}
where $\LL(\mathcal{H})$ denotes the set of linear operators on $\mathcal{H}$.  In gauge theories it is therefore quite natural to interpret the decomposition \eqref{hilbertdecomp} as being induced by the algebra  $\mathcal{A}(r)$ of gauge-invariant operators in the region $r$ and its commutant $\mathcal{A}(\ol{r})$, the algebra of gauge-invariant operators in $\ol{r}$ \cite{Casini:2013rba,Harlow:2016vwg}. The degrees of freedom labeled by $\alpha$ are precisely the shared center of these two algebras.

We now turn to gravity.  In general relativity the gauge transformations are diffeomorphisms, and infinitesimally they act on any tensor field $\phi$ via the Lie derivative
\be
\delta_\xi \phi=\mathcal{L}_\xi \phi.
\ee
At leading order in $G$ it will be sufficient to study pure general relativity with a negative cosmological constant, with action
\be
S=\frac{1}{16\pi G}\int_M d^dx \sqrt{-g}\big(\cR+(d-2)(d-1)\big)+S_{bd}.
\ee
Here $S_{bd}$ is a set of boundary terms which live at $\partial M$, which in addition to the Gibbons-Hawking term may include additional ``holographic renormalization'' terms depending on the boundary induced metric.  The symplectic form $\Omega$ for general relativity with this action can be constructed using standard techniques \cite{Iyer:1994ys}, and is invariant under gauge transformations $\xi$ which vanish at $\partial M$.  We however would like to define a phase space for a spatial subregion $r\subset \Sigma$ in gravity.  Using the machinery of \cite{Iyer:1994ys} it is not difficult to show that under a gauge transformation $\delta_\xi g=\mathcal{L}_\xi g$ for which the generating vector field $\xi^\mu$ vanishes at $\partial M$, we have
\be\label{presympR}
\Omega_{r}(\mathcal{L}_\xi g,\delta g)=-\int_{\hat{\partial r}}\left(\delta Q_\xi-\xi\cdot \theta\right),
\ee
where
\be
\xi\cdot \theta\equiv \frac{1}{16\pi G}\xi_\mu\left(g^{\mu\alpha}\nabla^\beta-g^{\alpha\beta}\nabla^\mu\right)\delta g_{\alpha\beta}\epsilon,
\ee
with $\epsilon$ being the $\epsilon$-tensor, and
\be
Q_\xi\equiv -\frac{1}{8\pi G} \star d \xi,
\ee
with $\xi$ now viewed as a one-form.  $\hat{\partial r}$ again denotes the part of $\partial r$ which does not intersect $\partial \Sigma$.  Thus $\Omega_r$ will again not be invariant under all the gauge transformations which we quotient by in the description of the full spacetime, so any construction of a phase space for ``just the degrees of freedom in $r$'' will need to address this.  The first obvious guess is to only quotient by gauge transformations $\xi^\mu$ which vanish at $\partial r$ and only consider variations which preserve the induced metric on $\partial r$.  This however is not quite sufficient: it gets rid of the second term in equation \eqref{presympR}, but the first still gives a nontrivial result\footnote{\label{boost}If we do allow the induced metric on $\hat{\partial r}$ to vary then there is also a term
\be
\frac{1}{16\pi G}\int_{\hat{\partial r}}\epsilon^{A}_{\phantom{A}B}\nabla_A\xi^B\delta \epsilon_{\hat{\partial r}},
\ee
which shows that $A[\hat{\partial r}]/(4G)$ is the generator of boosts in the normal plane around $\hat{\partial r}$.}
\be
\Omega_r(\mathcal{L}_\xi g,\delta g)=\frac{1}{16\pi G}\int_{\hat{\partial r}}\epsilon_{\hat{\partial r}}\partial_A\xi^B\delta \epsilon^A_{\phantom {A}B},
\ee
where $\epsilon_{\hat{\partial r}}$ is the $\epsilon$ tensor on $\hat{\partial r}$ and $\epsilon_{AB}$ is the $\epsilon$ tensor in the plane normal to $\hat{\partial r}$.  The quantity $\epsilon^A_{\phantom {A}B}$ depends on the metric only through the conformal structure of the two-dimensional metric in that normal plane, so this term will only vanish if we further restrict the metric variations $\delta g$ by requiring them to preserve that conformal structure.  This then requires us to further restrict the $\xi^\mu$ we quotient by to be conformal killing vectors in the normal plane \cite{Donnelly:2016auv}.  Thus we again define subregion phase spaces $\mathcal{P}_r(\alpha)$ and $\mathcal{P}_{\ol{r}}(\alpha)$, where $\alpha$ denotes an induced metric on $\hat{\partial r}$ and a conformal structure in the normal plane, and we have quotiented only by diffeomorphisms which vanish at $\hat{\partial r}$ and are conformal Killing vectors in the normal plane. We can then introduce a phase space on the full Cauchy slice as in equation \eqref{phasedecomp}, where again we sum over one element of each equivalence class under diffeomorphisms which do not necessarily vanish at $\hat{\partial r}$ and are not required to be conformal Killing vectors in the normal plane of the induced metrics on $\hat{\partial r}$ and conformal structures in the normal plane.  Since all conformal structures are gauge-equivalent on a two-dimensional plane, this means that the set $S$ will be equivalent just to the set of induced metrics on $\hat{\partial r}$ modulo diffeomorphisms there.  As in electromagnetism this will not quite be the full phase space of general relativity on this slice, since the canonical conjugates of the $\alpha$ are missing, but the two phase spaces will again lead to the same quantum theory with a Hilbert space decomposition \eqref{hilbertdecomp}.

The application of this result which is of interest for us in AdS/CFT is to define the gravitational dynamics within the entanglement wedge $W_R$ of a boundary subregion $R$.  We remind the reader that for any boundary spatial subregion $R$ the HRT surface is defined as the codimension-two achronal surface $\gamma_R$ obeying the following criteria:
\bi
\item $\partial \gamma_R=\partial R$.
\item The area of $\gamma_R$ is extremal under variations which preserve the previous condition.
\item There exists an achronal surface $\Sigma_R$ such that $\partial \Sigma_R=\gamma_R\cup R$.
\item If there is more than one surface obeying the previous criteria, we pick the one of smallest area.
\ei
The entanglement wedge $W_R$ of $R$ is then defined as the bulk domain of dependence of any such $\Sigma_R$.  In the above construction we should then take $r=\Sigma_R$ and $\hat{\partial r}=\gamma_R$ to arrive at a phase space construction of entanglement wedge dynamics.  The ``central'' degrees of freedom $S$ then consist of the induced metric on the HRT surface $\gamma_R$ modulo diffeomorphisms, which in particular includes its area: this confirms the argument of \cite{Harlow:2016vwg} that the ``area operator'' in the quantum Ryu-Takayanagi formula should be thought of as being in the center of the algebra of operators in the entanglement wedge.

\subsection{Computing boundary Renyi entropy at fixed area}\label{renyisec}

\bfig
\includegraphics[height=4cm]{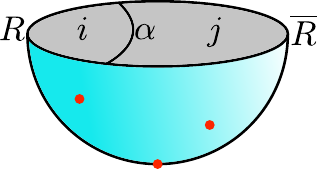}
\caption{Using the bulk Euclidean path integral to prepare a state.  The red dots are boundary sources we can adjust to change which state we prepare, and the center degrees of freedom $\alpha$ are the induced metric data on the HRT surface $\gamma_R$.}\label{statesfig}
\efig
Having constructed a classical phase space for gravity in the entanglement wedge, we may now quantize it, and in particular we may do so using the path integral.  For simplicity we restrict to states which possess a moment of time-reflection symmetry and can be prepared by a Euclidean path integral.\footnote{All of our arguments should be convertible to Lorentzian arguments along the lines of those in \cite{Dong:2016hjy}.}  These path integrals will prepare states in a Hilbert space with the structure \eqref{hilbertdecomp}, so they will have the form
\be\label{genstate}
|\psi\ran=\sum_{\alpha,i,j}C_{\alpha,ij}|\alpha,i\ran_R|\alpha,j\ran_{\ol{R}}.
\ee
We give a graphical illustration of such a state in figure \ref{statesfig}.  This representation is convenient because it allows to understand how projections onto definite values for $\alpha$ can be inserted into the cutting and gluing of gravitational path integrals.  We now use this to compute the boundary Renyi entropies for states which have been projected in this way.

In particular, let us consider a new state obtained by projecting the state $|\y\>$ in Eq.~\er{genstate} to a fixed area $\hat{A}$ on the HRT surface $\g_R$.\footnote{One can also consider states where we fix the entire induced metric on $\gamma_R$; we discuss these at the end of this subsection.}  This fixed-area state $|\y_{\hat{A}}\>$ can be thought of as an eigenstate of the ``area operator'' in the quantum RT formula.  Explicitly, it has the same form as in Eq.~\er{genstate} with the sum restricted to a subset of $\a$ for which the total area $A[\a]$ of $\g_R$ is $\hat{A}$:
\be\label{fixedA}
|\y_{\hat{A}}\>=\sum_{\alpha,i,j: A[\a]=\hat{A}}C_{\alpha,ij}|\alpha,i\ran_R|\alpha,j\ran_{\ol{R}}.
\ee
This state is prepared by the same bulk path integral with boundary sources that prepares $|\y\>$, but with the extra constraint that only configurations where the area of $\g_R$ is $\hat{A}$ are integrated over.

The norm of such a fixed-area state is calculated by a ``full'' bulk path integral obtained by gluing the path integral preparing $|\y_{\hat{A}}\>$ with a conjugate path integral preparing $\<\y_{\hat{A}}|$.  The two path integrals are glued together along the achronal surfaces $\S_R$ and $\S_{\ol{R}}$, with the fixed-area constraint enforced on $\g_R$:
\be\la{norm}
\<\y_{\hat{A}}|\y_{\hat{A}}\> = \int \mathcal{D}g \Big|_{A_{\g_R}[g]=\hat{A}} e^{-I[g]} \eq Z_1
\ee
where $A_{\g_R}[g]$ is the area of $\g_R$ in the metric $g$, $I[g]$ is the bulk Euclidean action, and dependence on the boundary sources is implicit in the path integral.  We call this full path integral $Z_1$ in anticipation of the discussion below on Renyi entropies.

In the semiclassical approximation, the path integral~\er{norm} is dominated by a saddle-point solution $g_1^c$, but as we have fixed the area of $\g_R$, $g_1^c$ is only required to satisfy the equations of motion away from $\g_R$ and is allowed to develop a uniform conical defect on $\g_R$.\footnote{We will provide a more general argument at the end of this subsection (when we discuss states with the entire induced metric on $\g_R$ fixed) which does not rely on showing that the singularity on $\g_R$ is a conical defect.}  To see the conical defect, we choose to enforce the fixed-area constraint in Eq.~\er{norm} by introducing a Lagrange multiplier $\m$:
\be\la{normcb}
Z_1 = \int \mathcal{D}g\, d\m\, e^{-I[g]-i\m (A_{\g_R}[g]-\hat{A})}.
\ee
The Lagrange multiplier term can be interpreted as the action of a cosmic brane with tension $i\m$.  Even though the cosmic brane is fixed to be on the HRT surface $\g_R$ (whose extremality can be defined here by the vanishing trace of extrinsic curvature as we approach $\g_R$), at the level of the saddle-point solution we could equivalently allow the location of the cosmic brane to be arbitrary (subject to the homology constraint) and require the total action in Eq.~\er{normcb} to be stationary with respect to variations of the brane location.  From this we find that the saddle-point geometry $g_1^c$ should satisfy the equations of motion everywhere including on the cosmic brane.  In the case of Einstein gravity that we focus on, codimension-2 cosmic branes backreact on the geometry by creating a conical deficit angle proportional to its tension~\cite{Vilenkin:1981zs}.\footnote{This follows from the observation in footnote~\ref{boost} that the area of $\g_R$ gives the generator of boosts in the normal plane around $\g_R$, and can also just be derived by studying the behavior of the Einstein equation in the vicinity of the brane.}  Therefore, the saddle-point solution of Eq.~\er{normcb} is characterized by a real conical defect geometry $g_1^c$ and a purely imaginary $\m=-i\m_E$ chosen so that the area of $\g_R$ agrees with $\hat{A}$ after the backreaction of the cosmic brane is taken into account.  We will refer to the conical opening angle in $g_1^c$ as $\p_1$.

\bfig
\includegraphics[height=4cm]{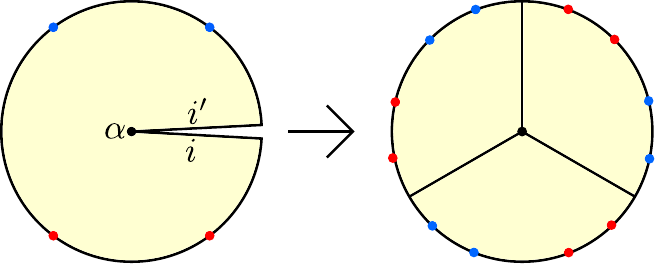}
\caption{Cutting open the $n=1$ gravitational path integral (left) and gluing together 3 copies to form a path integral for computing the third boundary Renyi entropy (right).  Red dots are the sources which produce the ``ket'' part of the density matrix, while blue dots are their CPT conjugates which produce the ``bra'' part.  The HRT surface $\gamma_R$ is represented by the black dot in the center, and is shared between all copies in the Renyi computation.}\label{renyigluefig}
\efig
Now let us calculate the boundary Renyi entropies on $R$ in the fixed-area state $|\y_{\hat{A}}\>$.  We will do this in two ways and obtain the same result.  The first way is a direct calculation of the Renyi entropies $S_n(\r_R)$ for integer $n\ge2$ by rewriting Eq.~\er{sndef} as
\be\la{snz}
S_n(\r_R) = \fr{1}{1-n} \log \fr{Z_n}{Z_1^n}
\ee
where $Z_n$ is an $n$-fold bulk path integral obtained by gluing together $n$ copies of the path integral preparing $|\y_{\hat{A}}\>$ (denoted by $P_1$, $\cdots$, $P_n$) and $n$ copies of the conjugate path integral preparing $|\y_{\hat{A}}\>$ (denoted by $\ol{P}_1$, $\cdots$, $\ol{P}_n$).  The manner of gluing is as follows: we first glue $P_i$ with $\ol{P}_i$ along $\S_{\ol{R}}$, and then glue $P_i$ with $\ol{P}_{i-1}$ along $\S_R$ (with the understanding that $\ol{P}_0$ means $\ol{P}_n$).  The common surface $\g_R$ in the resulting path integral is constrained to have area $\hat{A}$ since each of the pieces we glue does.   A natural bulk interpretation of this gluing procedure is that the first step of gluing $P_i$ with $\ol{P}_i$ along $\S_{\ol{R}}$ generates $n$ copies of an unnormalized density matrix which can be viewed as the bulk version of $\r_R$ and may also be obtained by cutting the path integral $Z_1$ open along $\S_R$.  The second step then forms $\Tr \r_R^n$ by cyclically gluing these density matrices together along $\S_R$.  We illustrate this in figure \ref{renyigluefig}.

Just like $Z_1$, the $n$-fold path integral $Z_n$ is given at leading order in the semiclassical approximation by the action of the dominant saddle-point solution $g_n^c$,
\be
Z_n = e^{-I[g_n^c]},
\ee
which can be used to rewrite Eq.~\er{snz} as
\be\la{snib}
S_n(\r_R) = \fr{I[g_n^c]-nI[g_1^c]}{n-1}.
\ee
A nice consequence of fixing the area of the HRT surface $\g_R$ is that the saddle-point geometry $g_n^c$ is extremely simple -- it is obtained by cyclically gluing $n$ identical copies of $g_1^c$ along $\S_R$ (after first cutting each one open along $\S_R$).  The resulting geometry $g_n^c$ is locally identical to $g_1^c$ away from $\g_R$ and has a conical defect on $\g_R$ with opening angle $\p_n=n\p_1$.

The action of $g_n^c$ (including $g_1^c$) consists of two contributions:
\be\la{ign}
I[g_n^c] = nI_{\text{away}}[g_1^c] + \fr{(n\p_1-2\pi) \hat{A}}{8\pi G}.
\ee
Here the first term comes from everywhere away from $\g_R$ and is therefore proportional to $n$.  The second term comes from a localized contribution on the conical defect $\g_R$ due to the fact that the Ricci scalar $\cR$ contains a delta function at $\g_R$ times a coefficient $2(2\pi-n\p_1)$ that is twice the conical deficit angle.  Plugging this into the Euclidean Einstein-Hilbert action $I=-\int d^dx \sqrt{g}\, \cR/(16\pi G)$, we find the second term in Eq.~\er{ign} where we have used the fixed area $\hat{A}$ of $\g_R$.

Plugging the action~\er{ign} into Eq.~\er{snib} we find that terms linear in $n$ cancel out, giving the Renyi entropy
\be\la{snflat}
S_n(\r_R) = \fr{\hat{A}}{4G}.
\ee
We have obtained this result for integer $n\ge2$, but it can trivially be analytically continued to an arbitrary $n$, leading to Renyi entropies that do not depend on $n$.  This suggests that the fixed-area state $|\y_{\hat{A}}\>$ has a flat entanglement spectrum on the boundary subregion $R$ at leading order in gravitational perturbation theory.

Now we provide an alternative way of obtaining the same boundary Renyi entropies.  It is simpler than the brute-force method used above, and has the advantage of giving the Renyi entropies for arbitrary $n$ directly without analytic continuation.  We start with the cosmic brane prescription for Renyi entropies derived in Refs.~\cite{Lewkowycz:2013nqa,Dong:2016fnf}.  It says that a refined version of the Renyi entropy defined by
\be\la{tsn}
\wtd{S}_n(\r_R) \eq n^2 \pa_n \(\fr{n-1}{n} S_n(\r_R)\) = -n^2 \pa_n \(\fr{1}{n} \log \Tr \r_R^n\)
\ee
is given by the area of a codimension-2 cosmic brane $\g_{R,n}$ inserted into the bulk solution:
\be\la{tsnb}
\wtd{S}_n(\r_R) = \fr{A[\g_{R,n}]}{4 G}.
\ee
The cosmic brane $\g_{R,n}$ is homologous to the subregion $R$ and creates a conical defect with deficit angle $2\pi(n-1)/n$ due to its tension $(n-1)/(4n G)$.  The refined Renyi entropy $\wtd{S}_n(\r_R)$ can be defined alternatively as the von Neumann entropy of the density matrix $\r_R^n / \Tr \r_R^n$.  The prescription~\er{tsnb} is a natural generalization of the RT formula with the cosmic brane $\g_{R,n}$ replacing the HRT surface $\g_R$.  Once we know the refined Renyi entropies $\wtd{S}_n$, the Renyi entropy $S_n$ for any $n$ is easily obtained by integrating~\er{tsn}:
\be\la{sni}
S_n(\r_R) = \fr{n}{n-1} \int_1^n \fr{\wtd{S}_{n'}(\r_R)}{n'^2} dn'.
\ee

Let us now apply this prescription to the fixed-area state $|\y_{\hat{A}}\>$.  We need to insert a cosmic brane $\g_{R,n}$ into the saddle-point geometry $g_1^c$ dominating the state norm $Z_1$.  It is worth noting that this cosmic brane $\g_{R,n}$ is different from (and introduced in addition to) the cosmic brane introduced earlier by the Lagrange multiplier in Eq.~\er{normcb} (which is already present in $g_1^c$).  However, the two cosmic branes coincide with each other and create a combined conical angle that is actually the same as the one in $g_1^c$ -- another great simplification due to fixing the area of the HRT surface $\g_R$.  To see this, we observe that if we place the cosmic brane $\g_{R,n}$ at exactly the location of $\g_R$ and do not change the geometry $g_1^c$ in any way, it would automatically satisfy the equations of motion away from $\g_R$ and have the same area $\hat{A}$ on $\g_R$.  Thus inserting the ``extra'' cosmic brane $\g_{R,n}$ does not affect the geometry at all -- the only thing that gets changed is the saddle point of the Lagrange multiplier, or equivalently the tension of the ``original'' cosmic brane already present in $g_1^c$: its saddle-point value in the path integral~\er{normcb} changes from $\m_E$ to $\m_E-(n-1)/(4nG)$ to ``absorb'' the tension of the extra cosmic brane $\g_{R,n}$ in order to keep the geometry unchanged and the equations of motion satisfied.

Therefore, the area of the cosmic brane $\g_{R,n}$ is $\hat{A}$ and Eq.~\er{tsnb} immediately leads to a constant refined Renyi entropy
\be\label{tsnflat}
\wtd{S}_n(\r_R) = \fr{\hat{A}}{4 G}.
\ee
Integrating this in Eq.~\er{sni} we reproduce the $n$-independent Renyi entropy~\er{snflat}.

In the discussion above we have focused on fixed-area states~\er{fixedA}, but we now point out that the same $n$-independent Renyi entropy applies to states obtained by projecting $|\y\>$ in Eq.~\er{genstate} to a fixed value $\hat\a$ of $\a$:
\be
|\y_{\hat\a}\>=\sum_{i,j}C_{\hat\alpha,ij}|\hat\alpha,i\ran_R|\hat\alpha,j\ran_{\ol{R}},
\ee
as long as the norm of the state is dominated semiclassically by a saddle-point geometry.  This fixed-$\a$ state is prepared by the same bulk path integral that prepares $|\y\>$ but with the extra constraint that the entire induced metric on the HRT surface $\g_R$ is given by $\hat\a$.

The norm of $|\y_{\hat\a}\>$ is again calculated by a full bulk path integral $Z_1$ defined in a way similar to Eq.~\er{norm}.  We assume that the path integral $Z_1$ is dominated in the semiclassical approximation by some saddle-point solution $g_1^c$.\footnote{We leave it to future work to verify this assumption and find the precise form of the singularity on $\g_R$ in general situations.}  The difference with the fixed-area case is that here $g_1^c$ is allowed to develop a more general singularity than a uniform conical defect on $\g_R$.  A specific example of such a singularity is a conical defect whose conical angle varies along $\g_R$, but in general it could be any singularity whose contribution to the bulk Lagrangian is a distribution (such as a delta function) localized on $\g_R$.

The boundary Renyi entropies on $R$ in the fixed-$\a$ state can be calculated by the same two methods used above.  The second method of introducing a cosmic brane $\g_{R,n}$ to calculate the refined Renyi entropy immediately leads to the same conclusion that the cosmic brane $\g_{R,n}$ would be placed at exactly the location of the singularity in $g_1^c$ without modifying the geometry in any way.  This gives the same $n$-independent refined Renyi entropy~\er{tsnflat} with $\hat A$ replaced by $A[\hat\a]$, the total area calculated from the fixed induced metric $\hat\a$.  Therefore, the Renyi entropies are also $n$-independent and given by
\be\la{snalpha}
S_n(\r_R) = \fr{A[\hat\a]}{4G}.
\ee
We can reproduce this result using the first, more direct method as well.  The Renyi entropies for integer $n\ge2$ are calculated from Eqs.~\er{snz} and~\er{snib} with the saddle-point geometry $g_n^c$ now satisfying the fixed-$\a$ constraint.  Again $g_n^c$ is simply obtained by cyclically gluing $n$ copies of $g_1^c$ along $\S_R$.  To calculate the contribution to the action from the singularity on $\g_R$ in $g_1^c$ and $g_n^c$, it is useful to regularize the singularity in $g_1^c$ by replacing a small neighborhood of $\g_R$ with a smooth geometry and to construct $g_n^c$ by gluing $n$ copies of this regularized version of $g_1^c$, with the understanding that the thickness of the neighborhood of $\g_R$ will be taken to zero eventually.  This version of $g_n^c$ has a simple, uniform conical defect on $\g_R$ with opening angle $2\pi n$, and its action is
\be\la{igngen}
I[g_n^c] = n I[g_1^c] + \fr{(n-1)A[\hat\a]}{4G}
\ee
where the first term comes from everywhere away from $\g_R$ (which is smooth in the regularized $g_1^c$) and the second comes from the uniform conical defect.  Eq.~\er{igngen} is the generalized version of Eq.~\er{ign} that applies here.  Plugging it into Eq.~\er{snib} we find the $n$-independent Renyi entropy~\er{snalpha} as promised.

Finally, we point out that we do not expect our conclusion of $n$-independent Renyi entropies to be modified by higher-derivative corrections in the gravitational action.  In these cases, the RT formula~\er{leadingRT} and its generalization~\er{tsnb} to refined Renyi entropies are modified by replacing the area $A$ by some generalized notion of area $A_{\text{gen}}$~\cite{Dong:2013qoa,Camps:2013zua,Dong:2017xht}.  This generalized area is an integral over $\g_R$ of some combination of local geometric invariants and its form is completely determined by the bulk action.  We expect that a calculation very similar to the one performed above in Einstein gravity would lead to the same $n$-independent Renyi entropies as in Eqs.~\er{snflat} and~\er{snalpha} with $A$ replaced by $A_{\text{gen}}$, although we leave the details to future work.

\subsection{The origin of \textit{n}-dependence}
We can now easily understand how the $n$-dependence of the Renyi entropy in unprojected semiclassical states arises.  Consider the fixed-area replicated path integral
\be
Z_n(\hat{A})=\int \mathcal{D}g_n e^{-I[g_n]}\delta(A_{\gamma_R}(g_n)-\hat{A}),
\ee
where here $\mathcal{D}g_n$ means that we are integrating over metrics (and other bulk fields) which at the Euclidean AdS boundary have $n$ copies of the sources preparing the state, as in figure \ref{renyigluefig}.  The unprojected path integral is then obtained by
\be
Z_n\equiv \int d\hat{A} Z_n(\hat{A})
\ee
In the semiclassical limit we have
\be
Z_n(\hat{A}) \approx e^{-I[g_n^c(\hat{A})]},
\ee
so we are interested in semiclassically evaluating the integral
\be\label{Aint}
Z_n=\int d\hat{A} e^{-I[g_n^c(\hat{A})]}.
\ee
The saddle point for this integral will be precisely the value $A_n$ for which the opening angle $\phi_1$ is $2\pi/n$, since ultimately in doing the integral over $\hat{A}$ we are just doing the full gravitational path integral in a different order and the saddle point $g_n^c(A_n)$ must be smooth and obey the Einstein equation at $\gamma_R$.  Moreover using the cosmic brane method \cite{Lewkowycz:2013nqa,Dong:2016fnf}, we then immediately see that the refined Renyi entropies are given by
\be
\wt{S}_n=\frac{A_n}{4G}.
\ee
Since $A_n$ is now a nontrivial function of $n$, we see that $\wt{S}_n$, and thus $S_n$, will be as well.  From this point of view the entire $n$-dependence of the Renyi entropies arises from the shifting of the saddle point value of $\hat{A}$ in \eqref{Aint} as we change $n$.

This discussion generalizes in a simple way to fixed-$\alpha$ states as well, we can also write
\be\label{alphaint}
Z_n=\sum_{\hat\alpha} Z_n(\hat{\alpha}),
\ee
and since $Z_n(\hat{\alpha})$ again leads to $n$-independent Renyi entropies we can view the $n$-dependence of the full Renyi entropies as coming from the shifting of the saddle point in the ``sum'' over $\alpha$ as we change $n$.

\section{A quantum error-correction interpretation}\label{qecsec}
In AdS/CFT the above gravitational discussion is all happening in the bulk, and must be embedded into the dual CFT some way.  Most states in the CFT will contain large black holes which in particular have swallowed whatever region we might be interested in.  In \cite{Almheiri:2014lwa} it was emphasized that the quantum structure of perturbative general relativity should be understood in AdS/CFT as holding only in a \textit{code subspace}, $\Hc$, of the full CFT Hilbert space.  This subspace is not unique, since we may be willing to tolerate some black holes which are far away from whatever physics we are considering, and if we are ambitious we may even try to include the microstates of some black holes within the code subspace degrees of freedom \cite{Harlow:2016vwg,Bao:2017guc,Hayden:2018khn}.  We may also want to exclude bulk degrees of freedom such as heavy matter fields which are describable within effective field theory. In general one can think of the choice of code subspace as being similar to the choice of a renormalization scheme: we choose it as is convenient for the particular problem we have in mind.  In fact this is more than analogy: doing renormalization group flow in the bulk is one example of changing our choice of code subspace.  For this paper we will take our code subspace to be the linear span of the set of states which can be prepared by a bulk Euclidean path integral with an $O(G^0)$ number of boundary sources of low scaling dimension.

Any state in this code subspace should obey the quantum Ryu-Takayanagi formula, which says that any bulk state $\rho$ prepared by a bulk path integral should have a dual CFT state $\wt{\rho}$ with the property that for any boundary subregion $R$ we have \cite{Faulkner:2013ana,Dong:2017xht}\footnote{From now on we view CFT states and operators as encoded quantities and denote them with tildes; this is slightly different from our notation in previous sections.  Note that these tildes are completely different from those in refined Renyi entropies $\wtd{S}_n$.}
\be \label{quantumRT}
S(\wt{\rho}_R)=\Tr\left(\rho \LL_R\right)+S(\rho_{r}).
\ee
Here $\LL_R$ is an operator localized on the HRT surface $\gamma_R$ which at leading order in $G$ is just $A[\gamma_R]/(4G)$, and $S(\rho_{r})$ is the entropy of $\rho$ restricted to the algebra of operators in the entanglement wedge $W_R$.  $\LL_R$ is required to be in the center of that algebra, and indeed this follows from the construction of section \ref{subregionsec} since the area of $\gamma_R$ is determined by its induced metric.  The von Neumann entropy of a state $\rho$ on a subalgebra $M$, which is a generalization of the definition for a subfactor, is given in terms of the diagonal blocks of $\rho$ in the decomposition \eqref{hilbertdecomp}.  Indeed we can represent each diagonal block of $\rho$ as $p_\alpha \rho_{r_\alpha \ol{r}_\alpha}$, with $\Tr\rho_{r_\alpha \ol{r}_\alpha}=1$ and $\sum_\alpha p_\alpha=1$, and the entropy is then (see \cite{Harlow:2016vwg} for more on this definition)
\be
S(\rho_r)\equiv -\sum_\alpha p_\alpha \log p_\alpha +\sum_\alpha p_\alpha S(\rho_{r_\alpha}).\label{algent}
\ee
One of the main results of \cite{Harlow:2016vwg} was that the quantum Ryu-Takayanagi formula \eqref{quantumRT} can hold in all states of a code subspace $\Hc\subset \mathcal{H}_R \otimes \mathcal{H}_{\ol{R}}$ if and only if the encoding map has a very specific form, which generalizes that of the quantum circuit shown in figure \ref{cutnetworkfig}.  Indeed if we algebraically decompose the code subspace $\Hc$ as in equation \eqref{hilbertdecomp}, we can choose a basis $|\wt{\alpha, i j}\ran$ as in figure \ref{statesfig}.  The result then is that the quantum RT formula requires $\HR$ and $\HRb$ to decompose as
\begin{align}\nonumber
\HR&=\oplus_\alpha\left(\mathcal{H}_{R_\alpha^1}\otimes \mathcal{H}_{R_\alpha^2}\right)\oplus \mathcal{H}_{R_3}\\
\la{hdecomp}
\HRb&=\oplus_\alpha\left(\mathcal{H}_{\ol{R}_\alpha^1}\otimes \mathcal{H}_{\ol{R}_\alpha^2}\right)\oplus \mathcal{H}_{\ol{R}_3},
\end{align}
where $\mathcal{H}_{R_\alpha^1}\cong \Hra$ and $\mathcal{H}_{\ol{R}_\alpha^1}\cong \Hrba$, and moreover that our complete basis for $\Hc$ must be obtainable as
\be\label{basisrep}
|\wt{\alpha, ij}\ran=U_R U_{\ol{R}}\left(|\alpha,i\ran_{R_\alpha^1}\otimes |\alpha,j\ran_{\ol{R}_\alpha^1}\otimes |\chi_\alpha\ran_{R^2_\alpha \ol{R}_\alpha^2}\right)
\ee
for some unitaries $U_R$, $U_{\ol{R}}$ on $\HR$ and $\HRb$ and some set of states $|\chi_\alpha\ran$.  In \cite{Harlow:2016vwg} codes with this structure were called ``operator algebra quantum error-correcting codes with complementary recovery'', since the structure \eqref{basisrep} also holds if and only if all operators in the $W_R$ algebra can be represented on $R$ and all operators in its commutant (the $W_{\ol{R}}$ algebra) can be represented on $\ol{R}$.  By taking the partial trace, \eqref{basisrep} immediately implies that for any bulk state $\rho$ on $\Hc$ we have
\be\label{rhorep}
\wt{\rho}_R=\sum_\alpha p_\alpha U_R\left(\rho_{R_\alpha^1}\otimes \chi_{R_\alpha^2}\right)U_R^\dagger,
\ee
where $\rho_{R_\alpha^1}$ has the same matrix elements as the state $\rho_{r_\alpha}$ appearing in the computation of the von Neumann entropy of the state $\rho$ on the algebra of operators in $W_R$ and $\chi_{R_\alpha^2}\equiv \Tr_{\ol{R}_\alpha^2}|\chi_\alpha\ran\lan\chi_\alpha|$.  Computing the von Neumann entropy of both sides of \eqref{rhorep} and using \eqref{algent}, we recover the quantum Ryu-Takayanagi formula \eqref{quantumRT} with the ``area operator'' given by
\be
\LL_R=\sum_\alpha S(\chi_{R_{\alpha}^2})I_{r_\alpha \ol{r}_\alpha}.
\ee
Showing that this is the only way for \eqref{quantumRT} to be satisfied is somewhat harder, and was the main content of \cite{Harlow:2016vwg}.

With equation \eqref{rhorep} in hand, it is a simple matter to consider also Renyi entropies. Indeed we have immediately that
\be
\Tr(\wt{\rho}_R^n)=\sum_\alpha p_\alpha^n \Tr(\rho_{r_\alpha}^n)\Tr(\chi_{R_\alpha^2}^n).
\ee
Just as in the quantum RT formula, in the semiclassical limit the ``bulk'' contribution from $\Tr(\rho_{r_\alpha}^n)$ will be subleading so the Renyi entropy for all states in the code subspace should obey
\be
\Tr(\wt{\rho}_R^n) \approx \sum_\alpha p_\alpha^n\Tr(\chi_{R_\alpha^2}^n).
\ee
Our proposal is that this equation is a CFT representation of equation \eqref{alphaint}.  Note in particular if we project onto a state of definite $\alpha$ then we are left just with $\Tr(\chi_{R_\alpha^2}^n)$, so we learn that the $n$-independence at fixed $\alpha$ we uncovered on the gravity side has a striking interpretation for holographic codes: the states $|\chi_\alpha\ran_{R_\alpha^2\ol{R}_\alpha^2}$ must have a flat entanglement spectrum on $R_\alpha^2$ for all $\alpha$.  For tensor network codes such as that in figure~\ref{cutnetworkfig} this was a consequence of the fact that this state was a tensor product of maximally entangled EPR pairs on the links of the network, but now we see it is a general property of holographic states at fixed $\alpha$.

We point out for completeness that the statements made above can also be seen from the refined Renyi entropies $\wtd{S}_n(\wtd{\r}_R)$ defined in Eq.~\er{tsn}.  To see this we now derive a general formula for $\wtd{S}_n(\wtd{\r}_R)$ that holds in any operator algebra quantum error-correcting code with complementary recovery and for any state in the code subspace.  We will see that it gives a nice code interpretation of the cosmic brane prescription~\er{tsnb} for refined Renyi entropies.  To do this we first note that $\wtd{S}_n(\wtd{\r}_R)$ is simply the von Neumann entropy of the density matrix $\wtd{\r}_{R,n} \eq \wtd{\r}_R^n/\Tr\wtd{\r}_R^n$ which by using Eq.~\er{rhorep} can be written as
\be
\wtd{\r}_{R,n} = \sum_\alpha p_{\alpha,n} U_R\left(\rho_{R_\alpha^1,n}\otimes \chi_{R_\alpha^2,n}\right)U_R^\dagger,\quad
p_{\a,n} \eq \fr{p_\a^n \Tr\,(\r_{R_\a^1}^n\otimes \c_{R_\a^2}^n)}{\sum_\b p_\b^n \Tr\,(\r_{R_\b^1}^n\otimes \c_{R_\b^2}^n)}
\ee
with $\r_{R_\alpha^1,n} \eq \r_{R_\a^1}^n / \Tr \r_{R_\a^1}^n$ and $\c_{R_\b^2,n}$ defined similarly.  Similar to the derivation of the quantum RT formula~\er{quantumRT} from Eq.~\er{rhorep}, we find its generalization to the refined Renyi entropies to be
\be\la{tsnqec}
\wtd{S}_n(\wtd{\r}_R) = \Tr \(\r \cL_{R,n}\) + \wtd{S}_n(\r_r)
\ee
with the ``Renyi area operator'' $\cL_{R,n}$ defined as
\be
\cL_{R,n} = \sum_\a \fr{p_{\a,n}}{p_\a} \wtd{S}_n(\c_{R_{\alpha}^2})I_{r_\alpha \ol{r}_\alpha}
\ee
and the refined Renyi entropy of $\r$ restricted to the algebra of operators in the entanglement wedge $W_R$ given by
\be
\wtd{S}_n(\rho_r)\equiv -\sum_\alpha p_{\a,n} \log p_{\a,n} +\sum_\alpha p_{\a,n} \wtd{S}_n(\rho_{r_\alpha}).
\ee
In the semiclassical limit, Eq.~\er{tsnqec} is dominated by the first term on the right hand side which becomes
\be\la{tsnlead}
\wtd{S}_n(\wtd{\r}_R) \approx \sum_\a p_{\a,n} \wtd{S}_n(\c_{R_\a^2}) = \fr{\sum_\a p_\a^n \Tr\,(\r_{R_\a^1}^n\otimes \c_{R_\a^2}^n)  \wtd{S}_n(\c_{R_\a^2})}{\sum_\b p_\b^n \Tr\,(\r_{R_\b^1}^n\otimes \c_{R_\b^2}^n)}.
\ee
Semiclassically, $\wtd{S}_n(\wtd{\r}_R)$ is given by $\wtd{S}_n(\c_{R_{\hat\a}^2})$ with the saddle-point value $\hat\a$ that dominates the sums over $\a$ of $n$-fold traces in Eq.~\er{tsnlead}.  This is (almost) precisely what the cosmic brane prescription~\er{tsnb} requires: the conical defect geometry created by the cosmic brane is exactly the $\bZ_n$ quotient of the (replica-symmetric) saddle-point geometry dominating the $n$-fold gravitational path integral [which is the bulk version of the $n$-fold traces in Eq.~\er{tsnlead}], with the same induced metric $\hat\a$ on $\g_R$ before or after the quotient.  Here we say ``almost'' because the gravitational prescription~\er{tsnb} gives the area of the conical defect which can be determined from its induced metric $\hat\a$ regardless of $n$, whereas Eq.~\er{tsnlead} gives $\wtd{S}_n(\c_{R_{\hat\a}^2})$ which appears to possibly depend on $n$.  The resolution of this puzzle is that $\wtd{S}_n(\c_{R_{\hat\a}^2})$ must in fact be independent of $n$ at least to leading order semiclassically!  We can also see this flat entanglement spectrum of $\c_{R_{\hat\a}^2}$ by comparing Eq.~\er{tsnlead} for a fixed-$\a$ state with $n$-independent Renyi entropies found on the gravity side.

There is a point about this connection which may at first seem confusing: our bulk path integral construction in figure \ref{renyigluefig} looks like we are computing the Renyi entropy only using the bulk degrees of freedom, but in the holographic interpretation that we have just given to the calculation we threw out the bulk contribution and the Renyi entropy came entirely from the states $|\chi_\alpha\ran$.  What happened?  The issue is that although the bulk construction in figure \ref{renyigluefig} resembles a bulk Renyi entropy calculation, strictly speaking it cannot be interpreted as such without a cutoff; otherwise the circle contracts and there is no trace interpretation in the bulk.  This is the same miracle by which the Euclidean path integral is able to compute the black hole entropy correctly without knowing the microscopic theory of quantum gravity.  In the CFT calculation we have those microstates in hand in the form of $|\chi_\alpha\ran$, and the calculation is dominated by them.  In the bulk calculation this UV information instead goes into the infrared value of Newton's constant $G$, which finds its way into the answer via the evaluation of the action on the semiclassical saddle point.

\section{An exponentiated JLMS formula?}\label{jlmssec}
There is an interesting interplay between the $n$-independence of Renyi entropies in fixed-$\alpha$ states and the JLMS formula relating bulk and boundary modular Hamiltonians.  To understand this relation, we first note that from \eqref{rhorep} we have
\be\la{kr}
\wt{K}_R^\rho\equiv -\log \wt{\rho}_R=-\sum_\alpha U_R\left(\log(p_\alpha \rho_{R_\alpha^1})\otimes I_{R_\alpha^2}+I_{R_\alpha^1}\otimes \log \chi_{R_\alpha^2}\right)U_R^\dagger.
\ee
Using the expression
\be
P_c\equiv \sum_\alpha U_RU_{\ol{R}}\left(I_{R_\alpha^1\ol{R}_\alpha^1}\otimes |\chi_\alpha\ran\lan\chi_\alpha|_{R_\alpha^2\ol{R}_\alpha^2}\right)U_{R}^\dagger U_{\ol{R}}^\dagger
\ee
for the projection operator onto $\Hc$, by direct calculation one can derive a version of the JLMS formula \cite{Jafferis:2015del,Harlow:2016vwg}
\be\label{jlms}
P_c\left(\wt{K}_R^\rho \otimes I_{\ol{R}}\right)P_c=\left(\wt{\LL}_R+\wt{K}_r^\rho\right)P_c,
\ee
where $\wt{\LL}_R$ is the encoded area operator
\be
\wt{\LL}_R\equiv \sum_\alpha U_RU_{\ol{R}}\left(S(\chi_{R_\alpha^2}) I_{R_\alpha^1\ol{R}_\alpha^1}\otimes |\chi_\alpha\ran\lan\chi_\alpha|_{R_\alpha^2\ol{R}_\alpha^2}\right)U_{R}^\dagger U_{\ol{R}}^\dagger
\ee
and $\wt{K}_r^\rho$ is the encoded bulk modular Hamiltonian
\be
\wt{K}_r^\rho\equiv \sum_\alpha U_RU_{\ol{R}}\left(-\log(p_\alpha \rho_{R_\alpha^1})\otimes I_{\ol{R}_\alpha^1}\otimes |\chi_\alpha\ran\lan\chi_\alpha|_{R_\alpha^2\ol{R}_\alpha^2}\right)U_{R}^\dagger U_{\ol{R}}^\dagger.
\ee

Now, given a modular Hamiltonian, it is natural to exponentiate it to generate modular flow \cite{Haag:1992hx}.  Since the JLMS formula \eqref{jlms} gives a relation between bulk and boundary modular Hamiltonians, one may ask if we can exponentiate it
into an equation of the form
\be\label{floweq}
P_c e^{is (\wt{K}_R^\rho\otimes I_{\ol{R}})}P_c=e^{is(\wt{\LL}_R+\wt{K}_r^\rho)}P_c,
\ee
which would say that we could compute the bulk modular evolution of a state in the code subspace using the boundary modular evolution.  Unfortunately however \eqref{floweq} only follows from \eqref{jlms} if $P_c$ and $\wt{K}_R^\rho \otimes I_{\ol R}$ are commuting quantum operators, and from a coding point of view there is no reason to expect this to be the case.  Indeed it is not hard to show that they will commute if and only if for all $\alpha$ we have
\be\label{chicond}
|\chi_\alpha\ran\lan \chi_\alpha|\log \chi_{R_\alpha^2}=\log \chi_{R_\alpha^2}|\chi_\alpha\ran\lan \chi_\alpha|.
\ee
Now however we learn something interesting: \eqref{chicond} holds if and only if the nonzero eigenvalues of $\chi_{R_\alpha^2}$ are all equal, which is precisely the structure we found at leading order in $G$ in section~\re{renyisec}!  In other words, the flat entanglement spectrum at fixed $\alpha$ which we found in gravity is closely related to the question of whether or not the exponentiated version \eqref{floweq} of the JLMS formula is valid, at least at leading order in $G$.

We can make this connection between the exponentiated JLMS formula and a flat entanglement spectrum at fixed $\alpha$ more explicit.  Indeed let $|\wt{\psi}_{\hat{\alpha}}\ran$ be some encoded state which has support only when $\alpha=\hat{\alpha}$. Using our formulae for $P_c$ and $\wt{K}_R^\rho$ we have
\be
\lan \wt{\psi}_{\hat{\alpha}}|e^{-is(\wt{K}_R^\rho\otimes I_{\ol{R}})}P_ce^{is(\wt{K}_R^\rho\otimes I_{\ol{R}})}|\wt{\psi}_{\hat{\alpha}}\ran=\Tr\left(\chi_{R_{\hat{\alpha}}^2}^{1+is}\right)\Tr\left(\chi_{R_{\hat{\alpha}}^2}^{1-is}\right).\label{flow2}
\ee
Equation \eqref{floweq} would imply that the left hand side of this equation is one, while on the right-hand side this in general would only be true for $s=0$.  The objects on the right-hand side are just the exponentials of the Renyi entropies of $\chi_{R_\alpha^2}$ analytically continued to complex $n$, so for equation \eqref{floweq} to hold through any particular order in $G$ we thus need these Renyi entropies to be $n$-independent to that order.\footnote{In testing this approximate flatness, the closeness of the right hand side of \eqref{flow2} to one is perhaps a better diagnostic of the accuracy of \eqref{floweq} than is \eqref{chicond}, since the former is a numerical relation and the latter is an operator equation.}  We suspect that our flatness result can be extended at least to $O(G^0)$ on the gravity side, but we leave a detailed study for future work.

In quantum error correction language, what we have learned is that the modular Hamiltonian $\wt{K}_R^\rho$ is a ``logical'' operator whose action preserves $\Hc$.  This is somewhat puzzling from the point of view of the CFT, where for generic regions $R$ it is only the ``full'' modular operator $\wt{K}_R^\rho\otimes I_{\ol{R}}-I_R\otimes \wt{K}_{\ol{R}}^\rho$ which preserves the set of low energy states.  And indeed we note that a calculation similar to that leading to \eqref{chicond} tells us that $[\wt{K}_R^\rho\otimes I_{\ol{R}}-I_R\otimes \wt{K}_{\ol{R}}^\rho,P_c]$ will vanish if and only if
\be
|\chi_\alpha\ran\lan \chi_\alpha|(\log \chi_{R_\alpha^2}-\log \chi_{\ol{R}_\alpha^2})+(\log \chi_{\ol{R}_\alpha^2}-\log \chi_{R_\alpha^2})|\chi_\alpha\ran\lan \chi_\alpha|=0,
\ee
which is true for \textit{any} $|\chi_\alpha\ran$ since the first and second terms vanish identically.  Thus the modular flow operation on operators generated by $\wt{K}_R^\rho\otimes I_{\ol{R}}-I_R\otimes \wt{K}_{\ol{R}}^\rho$ will always send logical operators to logical operators\footnote{An immediate corollary is that the modular flow generated by $\wt K^\rho_R\otimes I_{\ol{R}}$ alone will always preserve the set of logical operators supported on $R$, even if this flow does not preserve $\Hc$.}, as needed for the proposal of \cite{Faulkner:2017vdd}.  Nonetheless it seems that the stronger condition \eqref{floweq} also holds, at least to leading order in $G$.  How this is compatible with the CFT picture is a question we leave for future work.

\section{Discussion}
In this paper we argued that states which are prepared by Euclidean gravitational path integrals have the property that, if we project them onto eigenstates of fixed area for the HRT surface $\gamma_R$, we obtain a state whose Renyi entropies are independent of $n$ at leading order $O(1/G)$ in Newton's constant $G$.  We further argued that the same is true for states where we act with a projection onto a definite value of the entire induced metric on $\gamma_R$.  We then gave the latter result an interpretation within quantum error correction as a flat entanglement spectrum for the states $|\chi_\alpha\ran$ appearing in equation \eqref{basisrep}.  Finally we argued that this flatness leads to a somewhat surprising strengthening \eqref{floweq} of the relationship \cite{Jafferis:2015del,Harlow:2016vwg} between bulk and boundary modular Hamiltonians.

It would be interesting to develop a more detailed understanding of how our fixed-area states are realized in the CFT.  In specific simple cases we can give a rough picture, at least to leading order in the bulk Newton constant $G$.  Note for example that flat-spectrum states arise in any microcanonical ensemble, where all states in a given energy range $\Delta E = [E_1,E_2]$ enter with equal weight.  Starting with a thermofield-double state $|\psi \rangle$ on a pair of CFTs, taking the two CFTs to be respectively $R$ and $\bar R$, and projecting onto fixed HRT-area $A$ thus yields a state $|\psi\rangle_A$ that, up to normalization, seems likely to resemble the microcanonical double state  $|\psi\rangle_{\rm micro} = e^{-S/2}\sum_{E \in \Delta E} |E\rangle_R |E\rangle_L$ (see e.g. \cite{Marolf:2018ldl}), where $|E\rangle_{R,L}$ denote eigenstates of the right- and left-CFTs and with $\Delta E$ an appropriate small-but-not-exponentially-small range of energies centered on a Schwarzschild-AdS black hole of the desired area.   Indeed,  since the conical singularities of our fixed-area saddles make no contribution to the induced metric on the surface of time-symmetry\footnote{This interesting fact is critical to the idea that they prepare standard states, which in particular satisfy the usual Hamiltonian and momentum constraints.  It can be seen geometrically, or from the fact that the associated cosmic brane lies entirely inside the surface of time-symmetry, so that its stress tensor has no components normal to the surface.  Since the constraints are precisely the components of the bulk equations of motion that involve such normals, the brane stress tensor makes no contribution.},
comparing with the results of \cite{Marolf:2018ldl} one sees immediately that our fixed-area projection and the microcanonical double are described by semi-classical bulk saddles that coincide on this surface, and which thus define the same Lorentz-signature bulk solution.  The same is true for any Renyi copy of the states.  As a result, (Renyi) entropies and light fields in  $|\psi\rangle_{\rm micro}$ agree with those in the desired fixed-area state at the level of bulk classical solutions.

However, the two states $|\psi \rangle_A$ and $|\psi\rangle_{\rm micro}$ will differ beyond this order.  In particular, the fixed-area state will have additional contributions from certain terms obtained from $|\psi\rangle_{\rm micro}$ by acting separately on either side with unitaries that preserve the code subspace on which the HRT-area operator is defined.    Since the density of states is an increasing function of energy, such unitaries typically raise the total energy on either side.   In the bulk, such terms describe
copies of the microcanonical black hole (and in particular with the same HRT-area) with additional matter or gravitational waves on either side of the HRT surface. The restriction to unitaries that preserve the code subspace should allow only an $O(1)$ number of such terms within any small range of energies.  As a result, one expects the entanglement entropy of $|\psi\rangle_A$ to differ from that of $|\psi \rangle_{\rm micro}$ only at the level of $O(1)$ corrections.  Indeed, since each additional such term has the same entropy $S=A/4G$ as states in $\Delta E$, terms in which the energy has been raised by $\delta E \gg T$ are highly suppressed by the Boltzmann factor of the original thermofield-double state $|\psi \rangle$.  One may think of them as describing subleading saddles that contribute to $|\psi\rangle_A$.  And terms in which the energy has been raised only by $\delta E \lesssim T$ can be described by the same leading-order bulk saddle as $|\psi\rangle_{\rm micro}$, but with the associated state of bulk quantum fields differing by $O(1)$ excitations.  In this sense, fixed-area projections of the thermofield-double state $|\psi\rangle$ are just microcanonical double states at leading order in $G$.

Fixed HRT-area states defined by projecting a generic state $|\psi\rangle$  should thus be similarly close to microcanonical-double states defined by projecting $|\psi\rangle$ onto spectral intervals $\Delta K$ defined by its modular Hamiltonian $K$ on $R$.  However, when $\partial R = \partial \bar R$ is non-empty, one should understand these states to be somewhat singular as their UV structure clearly differs significantly from that of the CFT vacuum.  And as above small discrepancies will remain.  Of course, it should be possible to map the HRT-area operator (and thus its spectral projections) to the CFT using the methods of e.g. \cite{Faulkner:2017vdd} and hence to construct fixed HRT-area states directly in the CFT.  However it remains unclear to us whether the result will takes an elegant form, and in particular whether it admits a natural generalization to non-holographic CFTs.  These are important points to address in future investigations.

The flatness we found at leading order in $G$ in fixed-area states is a striking result.
Perhaps the most important task following up on this work is thus to use gravitational arguments to understand how far this flatness extends in the expansion in $G$.    This may also shed light on the meaning of the strengthened JLMS relation \eqref{floweq}.  Another interesting project would be to study in more detail how Renyi entropies behave in the tensor networks constructed in \cite{Donnelly:2016qqt} which have center degrees of freedom on the links.  It would also be interesting to establish similar results in higher-derivative theories of gravity, and it would be enlightening to work out the conical geometries in section \ref{renyisec} in some more detail in simple examples.  We are optimistic that the recent interplay between quantum gravity and quantum information theory has yet more to teach us.

\paragraph{Acknowledgments}
It is a pleasure to thank Tom Faulkner, Gary Horowitz, Ted Jacobson, Aitor Lewkowycz, Juan Maldacena, and Henry Maxfield for useful conversations.  We also thank Chris Akers and Pratik Rath for discussions on their related work \cite{Akers:2018fow} which is appearing jointly with this paper.  XD was supported in part by the National Science Foundation under Grant No.\ PHY-1820908.  DH and DM were supported in part by the Simons Foundation.  DH was also supported by US Department of Energy grants DE-SC0018944 and DE-SC0019127 and the MIT Department of Physics.   XD and DM were also supported in part by funds from the University of California.  This work was developed in part at the Aspen Center for Physics which is supported by the National Science Foundation under Grant No.\ PHY-1607611 and at the KITP which is supported in part by the National Science Foundation under Grant No.\ PHY-1748958.

\bibliographystyle{utcaps}
\bibliography{Renyi}

\providecommand{\href}[2]{#2}\begingroup\raggedright\begin{thebibliography}{10}

\bibitem{Sorkin:2014kta}
R.~D. Sorkin, ``{1983 paper on entanglement entropy: "On the Entropy of the
  Vacuum outside a Horizon"},'' in {\em {Proceedings, 10th International
  Conference on General Relativity and Gravitation: Padua, Italy, July 4-9,
  1983}}, vol.~2, pp.~734--736.
\newblock 1984.
\newblock
\href{http://arxiv.org/abs/1402.3589}{{\ttfamily arXiv:1402.3589 [gr-qc]}}.
\newblock

\bibitem{Bombelli:1986rw}
L.~Bombelli, R.~K. Koul, J.~Lee, and R.~D. Sorkin, ``{A Quantum Source of
  Entropy for Black Holes},''
\href{http://dx.doi.org/10.1103/PhysRevD.34.373}{{\em Phys. Rev.} {\bfseries
  D34} (1986) 373--383}.

\bibitem{Srednicki:1993im}
M.~Srednicki, ``{Entropy and area},''
  \href{http://dx.doi.org/10.1103/PhysRevLett.71.666}{{\em Phys. Rev. Lett.}
  {\bfseries 71} (1993) 666--669},
\href{http://arxiv.org/abs/hep-th/9303048}{{\ttfamily arXiv:hep-th/9303048
  [hep-th]}}.

\bibitem{Frolov:1993ym}
V.~P. Frolov and I.~Novikov, ``{Dynamical origin of the entropy of a black
  hole},'' \href{http://dx.doi.org/10.1103/PhysRevD.48.4545}{{\em Phys. Rev.}
  {\bfseries D48} (1993) 4545--4551},
\href{http://arxiv.org/abs/gr-qc/9309001}{{\ttfamily arXiv:gr-qc/9309001
  [gr-qc]}}.

\bibitem{Maldacena:2001kr}
J.~M. Maldacena, ``{Eternal black holes in anti-de Sitter},''
  \href{http://dx.doi.org/10.1088/1126-6708/2003/04/021}{{\em JHEP} {\bfseries
  04} (2003) 021},
\href{http://arxiv.org/abs/hep-th/0106112}{{\ttfamily arXiv:hep-th/0106112
  [hep-th]}}.

\bibitem{Ryu:2006bv}
S.~Ryu and T.~Takayanagi, ``{Holographic derivation of entanglement entropy
  from AdS/CFT},'' \href{http://dx.doi.org/10.1103/PhysRevLett.96.181602}{{\em
  Phys. Rev. Lett.} {\bfseries 96} (2006) 181602},
\href{http://arxiv.org/abs/hep-th/0603001}{{\ttfamily arXiv:hep-th/0603001
  [hep-th]}}.

\bibitem{Ryu:2006ef}
S.~Ryu and T.~Takayanagi, ``{Aspects of Holographic Entanglement Entropy},''
  \href{http://dx.doi.org/10.1088/1126-6708/2006/08/045}{{\em JHEP} {\bfseries
  08} (2006) 045},
\href{http://arxiv.org/abs/hep-th/0605073}{{\ttfamily arXiv:hep-th/0605073
  [hep-th]}}.

\bibitem{Hubeny:2007xt}
V.~E. Hubeny, M.~Rangamani, and T.~Takayanagi, ``{A Covariant holographic
  entanglement entropy proposal},''
  \href{http://dx.doi.org/10.1088/1126-6708/2007/07/062}{{\em JHEP} {\bfseries
  07} (2007) 062},
\href{http://arxiv.org/abs/0705.0016}{{\ttfamily arXiv:0705.0016 [hep-th]}}.

\bibitem{Swingle:2009bg}
B.~Swingle, ``{Entanglement Renormalization and Holography},''
  \href{http://dx.doi.org/10.1103/PhysRevD.86.065007}{{\em Phys. Rev.}
  {\bfseries D86} (2012) 065007},
\href{http://arxiv.org/abs/0905.1317}{{\ttfamily arXiv:0905.1317
  [cond-mat.str-el]}}.

\bibitem{VanRaamsdonk:2010pw}
M.~Van~Raamsdonk, ``{Building up spacetime with quantum entanglement},''
  \href{http://dx.doi.org/10.1007/s10714-010-1034-0,
  10.1142/S0218271810018529}{{\em Gen. Rel. Grav.} {\bfseries 42} (2010)
  2323--2329}, \href{http://arxiv.org/abs/1005.3035}{{\ttfamily arXiv:1005.3035
  [hep-th]}}.
[Int. J. Mod. Phys.D19,2429(2010)].

\bibitem{Bianchi:2012ev}
E.~Bianchi and R.~C. Myers, ``{On the Architecture of Spacetime Geometry},''
  \href{http://dx.doi.org/10.1088/0264-9381/31/21/214002}{{\em Class. Quant.
  Grav.} {\bfseries 31} (2014) 214002},
\href{http://arxiv.org/abs/1212.5183}{{\ttfamily arXiv:1212.5183 [hep-th]}}.

\bibitem{Lewkowycz:2013nqa}
A.~Lewkowycz and J.~Maldacena, ``{Generalized gravitational entropy},''
  \href{http://dx.doi.org/10.1007/JHEP08(2013)090}{{\em JHEP} {\bfseries 08}
  (2013) 090},
\href{http://arxiv.org/abs/1304.4926}{{\ttfamily arXiv:1304.4926 [hep-th]}}.

\bibitem{Maldacena:2013xja}
J.~Maldacena and L.~Susskind, ``{Cool horizons for entangled black holes},''
  \href{http://dx.doi.org/10.1002/prop.201300020}{{\em Fortsch. Phys.}
  {\bfseries 61} (2013) 781--811},
\href{http://arxiv.org/abs/1306.0533}{{\ttfamily arXiv:1306.0533 [hep-th]}}.

\bibitem{Balasubramanian:2014hda}
V.~Balasubramanian, P.~Hayden, A.~Maloney, D.~Marolf, and S.~F. Ross,
  ``{Multiboundary Wormholes and Holographic Entanglement},''
  \href{http://dx.doi.org/10.1088/0264-9381/31/18/185015}{{\em Class. Quant.
  Grav.} {\bfseries 31} (2014) 185015},
\href{http://arxiv.org/abs/1406.2663}{{\ttfamily arXiv:1406.2663 [hep-th]}}.

\bibitem{Marolf:2015vma}
D.~Marolf, H.~Maxfield, A.~Peach, and S.~F. Ross, ``{Hot multiboundary
  wormholes from bipartite entanglement},''
  \href{http://dx.doi.org/10.1088/0264-9381/32/21/215006}{{\em Class. Quant.
  Grav.} {\bfseries 32} no.~21, (2015) 215006},
\href{http://arxiv.org/abs/1506.04128}{{\ttfamily arXiv:1506.04128 [hep-th]}}.

\bibitem{Freedman:2016zud}
M.~Freedman and M.~Headrick, ``{Bit threads and holographic entanglement},''
  \href{http://dx.doi.org/10.1007/s00220-016-2796-3}{{\em Commun. Math. Phys.}
  {\bfseries 352} no.~1, (2017) 407--438},
\href{http://arxiv.org/abs/1604.00354}{{\ttfamily arXiv:1604.00354 [hep-th]}}.

\bibitem{Almheiri:2014lwa}
A.~Almheiri, X.~Dong, and D.~Harlow, ``{Bulk Locality and Quantum Error
  Correction in AdS/CFT},''
  \href{http://dx.doi.org/10.1007/JHEP04(2015)163}{{\em JHEP} {\bfseries 04}
  (2015) 163},
\href{http://arxiv.org/abs/1411.7041}{{\ttfamily arXiv:1411.7041 [hep-th]}}.

\bibitem{Jafferis:2015del}
D.~L. Jafferis, A.~Lewkowycz, J.~Maldacena, and S.~J. Suh, ``{Relative entropy
  equals bulk relative entropy},''
  \href{http://dx.doi.org/10.1007/JHEP06(2016)004}{{\em JHEP} {\bfseries 06}
  (2016) 004},
\href{http://arxiv.org/abs/1512.06431}{{\ttfamily arXiv:1512.06431 [hep-th]}}.

\bibitem{Dong:2016eik}
X.~Dong, D.~Harlow, and A.~C. Wall, ``{Reconstruction of Bulk Operators within
  the Entanglement Wedge in Gauge-Gravity Duality},''
  \href{http://dx.doi.org/10.1103/PhysRevLett.117.021601}{{\em Phys. Rev.
  Lett.} {\bfseries 117} no.~2, (2016) 021601},
\href{http://arxiv.org/abs/1601.05416}{{\ttfamily arXiv:1601.05416 [hep-th]}}.

\bibitem{Harlow:2016vwg}
D.~Harlow, ``{The Ryu–Takayanagi Formula from Quantum Error Correction},''
  \href{http://dx.doi.org/10.1007/s00220-017-2904-z}{{\em Commun. Math. Phys.}
  {\bfseries 354} no.~3, (2017) 865--912},
\href{http://arxiv.org/abs/1607.03901}{{\ttfamily arXiv:1607.03901 [hep-th]}}.

\bibitem{Holzhey:1994we}
C.~Holzhey, F.~Larsen, and F.~Wilczek, ``{Geometric and renormalized entropy in
  conformal field theory},''
  \href{http://dx.doi.org/10.1016/0550-3213(94)90402-2}{{\em Nucl. Phys.}
  {\bfseries B424} (1994) 443--467},
\href{http://arxiv.org/abs/hep-th/9403108}{{\ttfamily arXiv:hep-th/9403108
  [hep-th]}}.

\bibitem{Calabrese:2004eu}
P.~Calabrese and J.~L. Cardy, ``{Entanglement entropy and quantum field
  theory},'' \href{http://dx.doi.org/10.1088/1742-5468/2004/06/P06002}{{\em J.
  Stat. Mech.} {\bfseries 0406} (2004) P06002},
\href{http://arxiv.org/abs/hep-th/0405152}{{\ttfamily arXiv:hep-th/0405152
  [hep-th]}}.

\bibitem{Headrick:2010zt}
M.~Headrick, ``{Entanglement Renyi entropies in holographic theories},''
  \href{http://dx.doi.org/10.1103/PhysRevD.82.126010}{{\em Phys. Rev.}
  {\bfseries D82} (2010) 126010},
\href{http://arxiv.org/abs/1006.0047}{{\ttfamily arXiv:1006.0047 [hep-th]}}.

\bibitem{Perlmutter:2013gua}
E.~Perlmutter, ``{A universal feature of CFT Rényi entropy},''
  \href{http://dx.doi.org/10.1007/JHEP03(2014)117}{{\em JHEP} {\bfseries 03}
  (2014) 117},
\href{http://arxiv.org/abs/1308.1083}{{\ttfamily arXiv:1308.1083 [hep-th]}}.

\bibitem{Perlmutter:2013paa}
E.~Perlmutter, ``{Comments on Renyi entropy in AdS$_3$/CFT$_2$},''
  \href{http://dx.doi.org/10.1007/JHEP05(2014)052}{{\em JHEP} {\bfseries 05}
  (2014) 052},
\href{http://arxiv.org/abs/1312.5740}{{\ttfamily arXiv:1312.5740 [hep-th]}}.

\bibitem{Dong:2016fnf}
X.~Dong, ``{The Gravity Dual of Renyi Entropy},''
  \href{http://dx.doi.org/10.1038/ncomms12472}{{\em Nature Commun.} {\bfseries
  7} (2016) 12472},
\href{http://arxiv.org/abs/1601.06788}{{\ttfamily arXiv:1601.06788 [hep-th]}}.

\bibitem{Dong:2018lsk}
X.~Dong, ``{Holographic Renyi Entropy at Large Energy Density},''
\href{http://arxiv.org/abs/1811.04081}{{\ttfamily arXiv:1811.04081 [hep-th]}}.

\bibitem{Harlow:2018fse}
D.~Harlow, ``{TASI Lectures on the Emergence of Bulk Physics in AdS/CFT},''
  \href{http://dx.doi.org/10.22323/1.305.0002}{{\em PoS} {\bfseries TASI2017}
  (2018) 002},
\href{http://arxiv.org/abs/1802.01040}{{\ttfamily arXiv:1802.01040 [hep-th]}}.

\bibitem{Pastawski:2015qua}
F.~Pastawski, B.~Yoshida, D.~Harlow, and J.~Preskill, ``{Holographic quantum
  error-correcting codes: Toy models for the bulk/boundary correspondence},''
  \href{http://dx.doi.org/10.1007/JHEP06(2015)149}{{\em JHEP} {\bfseries 06}
  (2015) 149},
\href{http://arxiv.org/abs/1503.06237}{{\ttfamily arXiv:1503.06237 [hep-th]}}.

\bibitem{Hayden:2016cfa}
P.~Hayden, S.~Nezami, X.-L. Qi, N.~Thomas, M.~Walter, and Z.~Yang,
  ``{Holographic duality from random tensor networks},''
  \href{http://dx.doi.org/10.1007/JHEP11(2016)009}{{\em JHEP} {\bfseries 11}
  (2016) 009},
\href{http://arxiv.org/abs/1601.01694}{{\ttfamily arXiv:1601.01694 [hep-th]}}.

\bibitem{Faulkner:2013ana}
T.~Faulkner, A.~Lewkowycz, and J.~Maldacena, ``{Quantum corrections to
  holographic entanglement entropy},''
  \href{http://dx.doi.org/10.1007/JHEP11(2013)074}{{\em JHEP} {\bfseries 11}
  (2013) 074},
\href{http://arxiv.org/abs/1307.2892}{{\ttfamily arXiv:1307.2892 [hep-th]}}.

\bibitem{Han:2017uco}
M.~Han and S.~Huang, ``{Discrete gravity on random tensor network and
  holographic Rényi entropy},''
  \href{http://dx.doi.org/10.1007/JHEP11(2017)148}{{\em JHEP} {\bfseries 11}
  (2017) 148},
\href{http://arxiv.org/abs/1705.01964}{{\ttfamily arXiv:1705.01964 [hep-th]}}.

\bibitem{Donnelly:2016qqt}
W.~Donnelly, B.~Michel, D.~Marolf, and J.~Wien, ``{Living on the Edge: A Toy
  Model for Holographic Reconstruction of Algebras with Centers},''
  \href{http://dx.doi.org/10.1007/JHEP04(2017)093}{{\em JHEP} {\bfseries 04}
  (2017) 093},
\href{http://arxiv.org/abs/1611.05841}{{\ttfamily arXiv:1611.05841 [hep-th]}}.

\bibitem{Cantcheff:2014zma}
M.~Botta~Cantcheff, ``{Area Operators in Holographic Quantum Gravity},''
\href{http://arxiv.org/abs/1404.3105}{{\ttfamily arXiv:1404.3105 [hep-th]}}.

\bibitem{Fursaev:2006ih}
D.~V. Fursaev, ``{Proof of the holographic formula for entanglement entropy},''
  \href{http://dx.doi.org/10.1088/1126-6708/2006/09/018}{{\em JHEP} {\bfseries
  09} (2006) 018},
\href{http://arxiv.org/abs/hep-th/0606184}{{\ttfamily arXiv:hep-th/0606184
  [hep-th]}}.

\bibitem{Faulkner:2017vdd}
T.~Faulkner and A.~Lewkowycz, ``{Bulk locality from modular flow},''
  \href{http://dx.doi.org/10.1007/JHEP07(2017)151}{{\em JHEP} {\bfseries 07}
  (2017) 151},
\href{http://arxiv.org/abs/1704.05464}{{\ttfamily arXiv:1704.05464 [hep-th]}}.

\bibitem{Donnelly:2016auv}
W.~Donnelly and L.~Freidel, ``{Local subsystems in gauge theory and gravity},''
  \href{http://dx.doi.org/10.1007/JHEP09(2016)102}{{\em JHEP} {\bfseries 09}
  (2016) 102},
\href{http://arxiv.org/abs/1601.04744}{{\ttfamily arXiv:1601.04744 [hep-th]}}.

\bibitem{Marolf:1994cp}
D.~Marolf, ``{Diffeomorphism invariant actions for partial systems},''
  \href{http://dx.doi.org/10.1016/0370-2693(95)00195-Q}{{\em Phys. Lett.}
  {\bfseries B349} (1995) 89--93},
\href{http://arxiv.org/abs/gr-qc/9411067}{{\ttfamily arXiv:gr-qc/9411067
  [gr-qc]}}.

\bibitem{Donnelly:2017jcd}
W.~Donnelly and S.~B. Giddings, ``{How is quantum information localized in
  gravity?},'' \href{http://dx.doi.org/10.1103/PhysRevD.96.086013}{{\em Phys.
  Rev.} {\bfseries D96} no.~8, (2017) 086013},
\href{http://arxiv.org/abs/1706.03104}{{\ttfamily arXiv:1706.03104 [hep-th]}}.

\bibitem{Donnelly:2018nbv}
W.~Donnelly and S.~B. Giddings, ``{Gravitational splitting at first order:
  Quantum information localization in gravity},''
  \href{http://dx.doi.org/10.1103/PhysRevD.98.086006}{{\em Phys. Rev.}
  {\bfseries D98} no.~8, (2018) 086006},
\href{http://arxiv.org/abs/1805.11095}{{\ttfamily arXiv:1805.11095 [hep-th]}}.

\bibitem{Speranza:2017gxd}
A.~J. Speranza, ``{Local phase space and edge modes for
  diffeomorphism-invariant theories},''
  \href{http://dx.doi.org/10.1007/JHEP02(2018)021}{{\em JHEP} {\bfseries 02}
  (2018) 021},
\href{http://arxiv.org/abs/1706.05061}{{\ttfamily arXiv:1706.05061 [hep-th]}}.

\bibitem{Kirklin:2018gcl}
J.~Kirklin, ``{Subregions, Minimal Surfaces, and Entropy in Semiclassical
  Gravity},''
\href{http://arxiv.org/abs/1805.12145}{{\ttfamily arXiv:1805.12145 [hep-th]}}.

\bibitem{Camps:2018wjf}
J.~Camps, ``{Superselection Sectors of Gravitational Subregions},''
\href{http://arxiv.org/abs/1810.01802}{{\ttfamily arXiv:1810.01802 [hep-th]}}.

\bibitem{Giddings:2019hjc}
S.~B. Giddings, ``{Gravitational dressing, soft charges, and perturbative
  gravitational splitting},''
\href{http://arxiv.org/abs/1903.06160}{{\ttfamily arXiv:1903.06160 [hep-th]}}.

\bibitem{Donnelly:2011hn}
W.~Donnelly, ``{Decomposition of entanglement entropy in lattice gauge
  theory},'' \href{http://dx.doi.org/10.1103/PhysRevD.85.085004}{{\em Phys.
  Rev.} {\bfseries D85} (2012) 085004},
\href{http://arxiv.org/abs/1109.0036}{{\ttfamily arXiv:1109.0036 [hep-th]}}.

\bibitem{Ghosh:2015iwa}
S.~Ghosh, R.~M. Soni, and S.~P. Trivedi, ``{On The Entanglement Entropy For
  Gauge Theories},'' \href{http://dx.doi.org/10.1007/JHEP09(2015)069}{{\em
  JHEP} {\bfseries 09} (2015) 069},
\href{http://arxiv.org/abs/1501.02593}{{\ttfamily arXiv:1501.02593 [hep-th]}}.

\bibitem{Soni:2015yga}
R.~M. Soni and S.~P. Trivedi, ``{Aspects of Entanglement Entropy for Gauge
  Theories},'' \href{http://dx.doi.org/10.1007/JHEP01(2016)136}{{\em JHEP}
  {\bfseries 01} (2016) 136},
\href{http://arxiv.org/abs/1510.07455}{{\ttfamily arXiv:1510.07455 [hep-th]}}.

\bibitem{Casini:2013rba}
H.~Casini, M.~Huerta, and J.~A. Rosabal, ``{Remarks on entanglement entropy for
  gauge fields},'' \href{http://dx.doi.org/10.1103/PhysRevD.89.085012}{{\em
  Phys. Rev.} {\bfseries D89} no.~8, (2014) 085012},
\href{http://arxiv.org/abs/1312.1183}{{\ttfamily arXiv:1312.1183 [hep-th]}}.

\bibitem{Iyer:1994ys}
V.~Iyer and R.~M. Wald, ``{Some properties of Noether charge and a proposal for
  dynamical black hole entropy},''
  \href{http://dx.doi.org/10.1103/PhysRevD.50.846}{{\em Phys. Rev.} {\bfseries
  D50} (1994) 846--864},
\href{http://arxiv.org/abs/gr-qc/9403028}{{\ttfamily arXiv:gr-qc/9403028
  [gr-qc]}}.

\bibitem{Dong:2016hjy}
X.~Dong, A.~Lewkowycz, and M.~Rangamani, ``{Deriving covariant holographic
  entanglement},'' \href{http://dx.doi.org/10.1007/JHEP11(2016)028}{{\em JHEP}
  {\bfseries 11} (2016) 028},
\href{http://arxiv.org/abs/1607.07506}{{\ttfamily arXiv:1607.07506 [hep-th]}}.

\bibitem{Vilenkin:1981zs}
A.~Vilenkin, ``{Gravitational Field of Vacuum Domain Walls and Strings},''
\href{http://dx.doi.org/10.1103/PhysRevD.23.852}{{\em Phys. Rev.} {\bfseries
  D23} (1981) 852--857}.

\bibitem{Dong:2013qoa}
X.~Dong, ``{Holographic Entanglement Entropy for General Higher Derivative
  Gravity},'' \href{http://dx.doi.org/10.1007/JHEP01(2014)044}{{\em JHEP}
  {\bfseries 01} (2014) 044},
\href{http://arxiv.org/abs/1310.5713}{{\ttfamily arXiv:1310.5713 [hep-th]}}.

\bibitem{Camps:2013zua}
J.~Camps, ``{Generalized entropy and higher derivative Gravity},''
  \href{http://dx.doi.org/10.1007/JHEP03(2014)070}{{\em JHEP} {\bfseries 03}
  (2014) 070},
\href{http://arxiv.org/abs/1310.6659}{{\ttfamily arXiv:1310.6659 [hep-th]}}.

\bibitem{Dong:2017xht}
X.~Dong and A.~Lewkowycz, ``{Entropy, Extremality, Euclidean Variations, and
  the Equations of Motion},''
  \href{http://dx.doi.org/10.1007/JHEP01(2018)081}{{\em JHEP} {\bfseries 01}
  (2018) 081},
\href{http://arxiv.org/abs/1705.08453}{{\ttfamily arXiv:1705.08453 [hep-th]}}.

\bibitem{Bao:2017guc}
N.~Bao and H.~Ooguri, ``{Distinguishability of black hole microstates},''
  \href{http://dx.doi.org/10.1103/PhysRevD.96.066017}{{\em Phys. Rev.}
  {\bfseries D96} no.~6, (2017) 066017},
\href{http://arxiv.org/abs/1705.07943}{{\ttfamily arXiv:1705.07943 [hep-th]}}.

\bibitem{Hayden:2018khn}
P.~Hayden and G.~Penington, ``{Learning the Alpha-bits of Black Holes},''
\href{http://arxiv.org/abs/1807.06041}{{\ttfamily arXiv:1807.06041 [hep-th]}}.

\bibitem{Haag:1992hx}
R.~Haag, {\em {Local quantum physics: Fields, particles, algebras}}.
\newblock
1992.
\newblock

\bibitem{Marolf:2018ldl}
D.~Marolf, ``{Microcanonical Path Integrals and the Holography of small Black
  Hole Interiors},''
\href{http://arxiv.org/abs/1808.00394}{{\ttfamily arXiv:1808.00394 [hep-th]}}.

\bibitem{Akers:2018fow}
C.~Akers and P.~Rath, ``{Holographic Renyi Entropy from Quantum Error
  Correction},'' \href{http://dx.doi.org/10.1007/JHEP05(2019)052}{{\em JHEP}
  {\bfseries 05} (2019) 052},
\href{http://arxiv.org/abs/1811.05171}{{\ttfamily arXiv:1811.05171 [hep-th]}}.

\end{thebibliography}\endgroup
\end{document}